\documentclass[12pt, a4paper]{article} 

\usepackage[english]{babel}
\usepackage[utf8]{inputenc}
\usepackage[T1]{fontenc}
\usepackage{lmodern}
\usepackage{microtype}
\usepackage{csquotes}

\usepackage{geometry}
\usepackage{setspace}
\usepackage{titlesec}
\titlespacing*{\section}{0pt}{1ex plus 0.5ex minus .2ex}{0.5ex plus .1ex}
\titlespacing*{\subsection}{0pt}{1ex plus 0.3ex minus .1ex}{0.5ex plus .1ex}

\usepackage{amsmath}
\usepackage{amssymb}
\usepackage{amsfonts}
\usepackage{amsthm}
\usepackage{siunitx}
\DeclareSIUnit{\molar}{M}
\DeclareSIUnit{\Molar}{M}

\usepackage{float} 

\usepackage{graphicx}
\usepackage{xcolor}
\definecolor{pastelblue}{rgb}{0.0, 0.5, 1.0}
\definecolor{redc}{rgb}{0.9,0.2,0.3}

\usepackage{tikz}
\usepackage{pgfplots}
\pgfplotsset{compat=1.17}
\usetikzlibrary{shapes, arrows, positioning}
\usepackage{booktabs}
\usepackage{multirow}
\usepackage{arydshln}
\usepackage{enumitem}

\usepackage[compatibility=false]{caption}
\usepackage{subcaption}
\DeclareCaptionLabelFormat{boldpipe}{\textbf{#1~#2|}\ }
\usepackage{authblk}
\usepackage{datetime}

\makeatletter
\renewcommand{\@makefnmark}{\textsuperscript{\fnsymbol{footnote}}}

\usepackage[
  backend=biber,
  natbib=true, 
  sorting=none,
  giveninits=true,
  uniquename=init,
  maxbibnames=3,  
  minbibnames=3,
  doi=true,
  url=false,
  isbn=false,
  eprint=false,
  date=year
]{biblatex}
\theoremstyle{definition}

\theoremstyle{remark}

\usepackage[colorlinks=true, citecolor=pastelblue, linkcolor=redc, urlcolor=redc]{hyperref}

\makeatother 

\title{Predicting MacroB12: Data-Driven Assessment of Pre-PEG B12 Utility and Clinical Caveats.}
\author[1,$\dagger$]{Carmen Frías-Ruiz}
\author[2,3,4,5,$\dagger$]{José María Gálvez-Navas}
\author[1]{Isabel García-Calcerrada}
\author[1]{Silvia Heras Flórez}
\author[6,*]{Rene Fabregas}

\affil[1]{\small Servicio de Análisis Clínicos, Hospital Universitario Nuestra Señora de Candelaria, Crta. Gral. del Rosario, 145, 38012, Santa Cruz de Tenerife, Spain}
\affil[2]{\small Centro de Investigación Biomédica en Red de Epidemiología y Salud Pública (CIBERESP), Spain}
\affil[3]{\small Andalusian School of Public Health, Cuesta del Observatorio, 4, Campus Universitario de Cartuja, 18011, Granada, Spain}
\affil[4]{\small Instituto de Investigación Biosanitaria ibs.GRANADA, Granada, Spain}
\affil[5]{\small Departamento de Bioquímica y Biología Molecular II, Facultad de Farmacia, Universidad de Granada, Campus Universitario de Cartuja, 18011, Granada, Spain}
\affil[6]{\small Departamento de Matemática Aplicada and Modeling Nature (MNat) Research Unit, Facultad de Ciencias, Universidad de Granada, 18071-Granada, Spain}
\affil[*]{\small Corresponding Author: Rene Fabregas, \href{mailto:rfabregas@ugr.es}{rfabregas@ugr.es}}
\affil[$\dagger$]{\small These authors contributed equally to this work.}

\date{\today} 

\begin{document}

\maketitle

\begin{abstract}
MacroB12 interference presents a significant challenge in the diagnosis of Vitamin B12 status, potentially masking true deficiency. To establish robust predictors and quantify the utility of pre-polyethene glycol (PEG) B12 levels, we conducted a retrospective analysis of 875 individuals with hypercobalaminemia (>1000~pg/mL), using multiple imputation to handle missing data. Multivariable regression modelling revealed that MacroB12 positivity (<30\% PEG recovery) was independently associated with a profile suggestive of autoimmunity, including older age (adjusted odds ratio [aOR]~1.03 per year, \textit{P}<0.001) and a Rheumatologic/Autoimmune diagnosis (aOR~2.96, \textit{P}=0.01). We also identified a counterintuitive haematological signature, with higher haemoglobin (aOR~1.14, \textit{P}=0.02) and Mean Corpuscular Volume (aOR~1.04, \textit{P}=0.01) predicting MacroB12. Receiver Operating Characteristic (ROC) analysis of pre-PEG B12 concentration yielded only moderate discriminatory power (Area Under the Curve [AUC]~=~0.744, 95\%~CI 0.707–0.782). The optimal threshold of 1584.0~pg/mL (sensitivity 71.3\%, specificity 69.7\%) serves to stratify clinical suspicion, but its performance confirms that it cannot replace definitive confirmatory testing. Our findings define a clinical and laboratory phenotype for suspected MacroB12 but highlight that confirmatory PEG precipitation remains essential for accurate diagnosis, particularly in older patients and those with autoimmune disease.

\vspace{1em} 
\end{abstract}


\section{Introduction}\label{sec:introduction}

Vitamin B12 (cobalamin), an essential water-soluble micronutrient, occupies a central position in fundamental cellular processes, including one-carbon metabolism critical for DNA synthesis and the metabolic pathways governing fatty acid and amino acid turnover \cite{DUNLOP1949754,  Obeid2019, NIHVitaminB12, Moravcova2025}. Unlike most vitamins, cobalamin synthesis is restricted to certain microorganisms, necessitating its exogenous acquisition by humans primarily through consumption of animal-derived foods where it accumulates \cite{Watanabe2007}. Within the circulation, B12 is chaperoned by two main proteins: transcobalamin and haptocorrin \cite{Markle1996, McCorvie2023}, with tight binding ensuring retention \cite{Fedosov2002,Fedosov2007} and supporting its diverse roles in neurological health and hematopoiesis \cite{Green2017,Thain2025}.

While clinical focus has historically centered on B12 deficiency \cite{Carmel2008}, the frequent occurrence of \textbf{hypercobalaminemia}—markedly elevated serum B12 levels—is a surprisingly frequent finding in routine laboratory practice, often exceeding deficiency rates \cite{Arendt2012}. This finding, increasingly recognized beyond benign explanations like supplementation, often signals significant underlying pathologies, including malignancies, hepato-renal disease, or autoimmune conditions, and has been independently associated with increased mortality risk \cite{Tal2010, Arendt2013, AmadoGarzon2024, Arendt2016, Eduin2023, Liu2024}. Nevertheless, standardized clinical algorithms for investigating high B12 remain notably absent \cite{Serraj2011, FloresGuerrero2019, FernandezLandazuri2024, SaeedParr2025}. A primary contributor to spuriously elevated B12 measurements is the presence of \textbf{MacroB12}, typically high-molecular-weight complexes of B12 with immunoglobulins (primarily IgG) that interfere with common automated immunoassays \cite{Olesen1968, Bowen2006, Delgado2023}. This \textbf{analytical interference} yields artifactually high serum concentrations, creating a significant diagnostic dilemma by potentially masking true functional B12 deficiency \cite{Remacha2014, Delgado2023}. Although reference methods like gel filtration chromatography exist, their complexity limits routine use \cite{LothianRoberts2016}. Polyethylene glycol (\textbf{PEG}) precipitation represents a widely accessible and pragmatic alternative \cite{Soleimani2019, OncelVanDemir2023,Yang2012}, yet distinguishing interference from true hypercobalaminemia remains challenging, hampered in part by a lack of universal consensus on optimal interpretative thresholds for PEG recovery \cite{Remacha2014, Soleimani2019}.

This long-standing analytical challenge has become increasingly prominent with the automation of clinical laboratories~\cite{Ermens2003, Arendt2013}; concurrently, the availability of large-scale clinical data and robust statistical tools provides a timely opportunity for its rigorous re-evaluation~\cite{Murdoch2013}. Prior investigations into MacroB12 have often been limited by smaller sample sizes, heterogeneous methodologies, or a failure to comprehensively adjust for the complex interplay of clinical confounders~\cite{Delgado2023}. Consequently, a clear understanding of the independent predictors distinguishing MacroB12 interference from other causes of hypercobalaminemia across diverse patient populations is lacking. Furthermore, while pre-PEG B12 levels are widely recognized as elevated in the presence of MacroB12, their quantitative discriminatory performance has not been rigorously established using contemporary statistical approaches, nor has an evidence-based optimal threshold been derived from large-scale clinical data to guide laboratory workflows and clinical suspicion. \textbf{This study directly addresses these critical knowledge gaps} by leveraging a large, representative clinical cohort (N=875) and employing robust statistical techniques, including multiple imputation and multivariable modeling. We hypothesized that specific clinical profiles, particularly autoimmune conditions, and distinct laboratory signatures beyond elevated pre-PEG B12 would independently associate with MacroB12 positivity. The \textbf{primary objectives} were therefore to: (1) robustly identify independent clinical and laboratory predictors of MacroB12 (defined by <30\% PEG recovery) using \emph{pooled multivariable regression models}; (2) quantitatively establish the diagnostic utility and data-driven optimal threshold of \textbf{pre-PEG B12 concentration} via \emph{Receiver Operating Characteristic (ROC) curve analysis}; and (3) provide foundational data to inform standardized diagnostic pathways for suspected MacroB12 interference, thereby enhancing diagnostic accuracy and potentially reducing unnecessary downstream investigations.

Herein, we provide a structured presentation of our findings. Section~\ref{sec:results} details the study cohort characteristics (Table~\ref{tab:baseline_char}) and presents key univariate comparisons (Fig.~\ref{fig:boxviolins}). We then detail findings from our multivariable regression models (logistic and Poisson) that identify independent predictors of MacroB12 status (Figs.~\ref{fig2:forest_plots} and \ref{fig3:forest_plots_pr}), and subsequently evaluate the discriminatory accuracy and optimal threshold for pre-PEG B12 using ROC curve analysis (Fig.~\ref{fig:roc_main}). Exploratory analyses, including factors associated with quantitative PEG recovery via \emph{Beta regression} and a \emph{Random Forest} assessment of variable importance, alongside ROC curves for secondary biomarkers and detailed methods for missing data handling and sensitivity analyses, are provided in the Supplementary Information (Supplementary Methods, Supplementary Results, Supplementary Table~S1, Supplementary Table~S2, Supplementary Fig.~S1, Supplementary Fig.~S2). The implications of these findings are contextualized within existing literature and clinical practice in Section~\ref{sec:discussion}, followed by concluding remarks in Section~\ref{sec:conclusions}. Comprehensive methodological details are provided in Section~\ref{sec:methods}.

\section{Results}\label{sec:results} 

\subsection{Cohort Demographics and Baseline MacroB12 Status}
The study cohort comprised 875 individuals referred for MacroB12 assessment, analyzed using multiply imputed datasets to account for missing data (Methods). The overall prevalence of MacroB12 positivity, defined by a standard post-polyethylene glycol (PEG) precipitation recovery of <30\%, was 27.1\% (n=237), with the remaining 72.9\% (n=638) classified as MacroB12 negative. A detailed comparison of baseline demographic, clinical, and laboratory characteristics between these two groups revealed significant differences (Table~\ref{tab:baseline_char}). Notably, patients with MacroB12 exhibited a distinct clinical profile: they were significantly older, with a median age difference exceeding a decade (Positive: 71.0 [IQR: 60.0--78.0] vs. Negative: 59.0 [42.0--73.0] years; \textit{P}<0.001). This age association aligns with previous observations suggesting an increased prevalence of autoantibodies or other interfering substances with age \cite{Herrmann2003}. Furthermore, MacroB12 positive individuals predominantly presented from ambulatory settings (76.4\% vs. 53.0\% for negatives), with correspondingly lower proportions from hospital or external center referrals (Table~\ref{tab:baseline_char}).
\begin{table}[ht!]
    \centering
    \caption{Baseline Characteristics of the Study Cohort, Stratified by MacroB12 Status.} 
\label{tab:baseline_char}
\resizebox{\textwidth}{!}{%
 \begin{tabular}{l c c c}
        \toprule
        Characteristic & Overall Cohort & Macro B12 Negative & Macro B12 Positive \\
         & (N = 875) & (N = 638) & (N = 237) \\
        \midrule
        \textbf{Demographics} & & & \\
        \quad Age (years) & 63 [47 – 76] & 59 [42 – 73] & 71 [60 – 79] \\
        \quad Sex, N (\%) & & & \\ 
        \qquad Male & 373 (42.6) & 284 (44.5) & 89 (37.6) \\
        \qquad Female & 502 (57.4) & 354 (55.5) & 148 (62.4) \\
        \addlinespace 
        \addlinespace
        \textbf{Clinical Context} & & & \\
        \quad Source, N (\%) & & & \\ 
        \qquad Ambulatory & 519 (59.3) & 338 (53.0) & 181 (76.4) \\
        \qquad Hospital & 234 (26.7) & 196 (30.7) & 38 (16.0) \\
        \qquad External Center & 122 (13.9) & 104 (16.3) & 18 (7.6) \\
        \quad \textbf{Diagnosis Present (N (\%))} & & & \\ 
        \qquad Diagnosis: Healthy/Unspecified & 410 (46.9) & 303 (47.5) & 107 (45.1) \\
        \qquad Diagnosis: Hematologic & 84 (9.6) & 76 (11.9) & 8 (3.4) \\
        \qquad Diagnosis: Hepatic & 73 (8.3) & 59 (9.2) & 14 (5.9) \\
        \qquad Diagnosis: Oncologic & 79 (9.0) & 57 (8.9) & 22 (9.3) \\
        \qquad Diagnosis: Rheum./Autoimmune & 57 (6.5) & 30 (4.7) & 27 (11.4) \\
        \qquad Diagnosis: Infectious & 23 (2.6) & 19 (3.0) & 4 (1.7) \\
        \qquad Diagnosis: Renal & 70 (8.0) & 54 (8.5) & 16 (6.8) \\
        \qquad Diagnosis: GI/Neo. Suspicion & 42 (4.8) & 31 (4.9) & 11 (4.6) \\
        \qquad Diagnosis: Hypertension & 82 (9.4) & 47 (7.4) & 35 (14.8) \\
        \qquad Diagnosis: Diabetes & 25 (2.9) & 12 (1.9) & 13 (5.5) \\
        \addlinespace 
        \textbf{Vitamin B12 Parameters} & & & \\
        \quad VB12 Pre-PEG (pg/mL)** & 1407 [1128 – 2000] & 1276 [1094 – 1714] & 2000 [1486 – 3810] \\
        \quad PEG Recovery (\%) & 39.7 [27.9 – 45.4] & 42.7 [38.3 – 47.7] & 15.5 [11.2 – 21.8] \\
        \addlinespace 
        \textbf{Concomitant Laboratory Markers} & & & \\
        \quad CRP (mg/L) (N valid=505) & 0.3 [0.1 – 3.0] & 0.6 [0.1 – 3.7] & 0.2 [0.1 – 1.0] \\
        \quad Serum Folate (ng/mL) (N valid=779) & 7.0 [4.8 – 11.1] & 6.9 [4.8 – 11.1] & 7.4 [4.9 – 11.1] \\
        \quad Hemoglobin (g/dL) & 13.3 [11.4 – 14.6] & 13.0 [10.9 – 14.4] & 13.9 [12.3 – 14.9] \\
        \quad MCV (fL) & 91.4 [88.0 – 95.3] & 90.9 [87.2 – 95.1] & 92.2 [89.5 – 96.2] \\
        \quad RDW (\%) & 13.8 [13.0 – 15.3] & 13.9 [13.0 – 15.6] & 13.6 [12.9 – 14.7] \\ 
        \bottomrule
    \end{tabular}
} 
\caption*{\footnotesize Data presented as Median [Interquartile Range] for continuous variables or N (\%) for categorical variables, based on pooled imputed dataset \#1 unless otherwise noted. MacroB12 Positive defined as PEG recovery <30\%. P-values comparing groups are presented in Fig.~\ref{fig:boxviolins} and statistical models. \textsuperscript{a}N valid=505 originally; \textsuperscript{b}N valid=779 originally. Includes 59 values reported as > limit in original data (handled via imputation). Abbreviations: CI, Confidence Interval; CRP, C-reactive protein; Ext, External; GI, Gastrointestinal; Hb, Hemoglobin; Hosp, Hospital; HTA, Hypertension; IQR, Interquartile Range; MCV, Mean Corpuscular Volume; Neo, Neoplastic; PEG, Polyethylene glycol; RDW, Red Cell Distribution Width; Rheum, Rheumatologic; Sens, Sensitivity; Spec, Specificity; Unspec., Unspecified; VB12, Vitamin B12.} 
\end{table}

\subsection{Univariate Associations Reveal a Paradoxical MacroB12 Signature} \label{sec:univariate_associations}

\begin{figure}[ht!]
    \centering
     \includegraphics[width=1\textwidth]{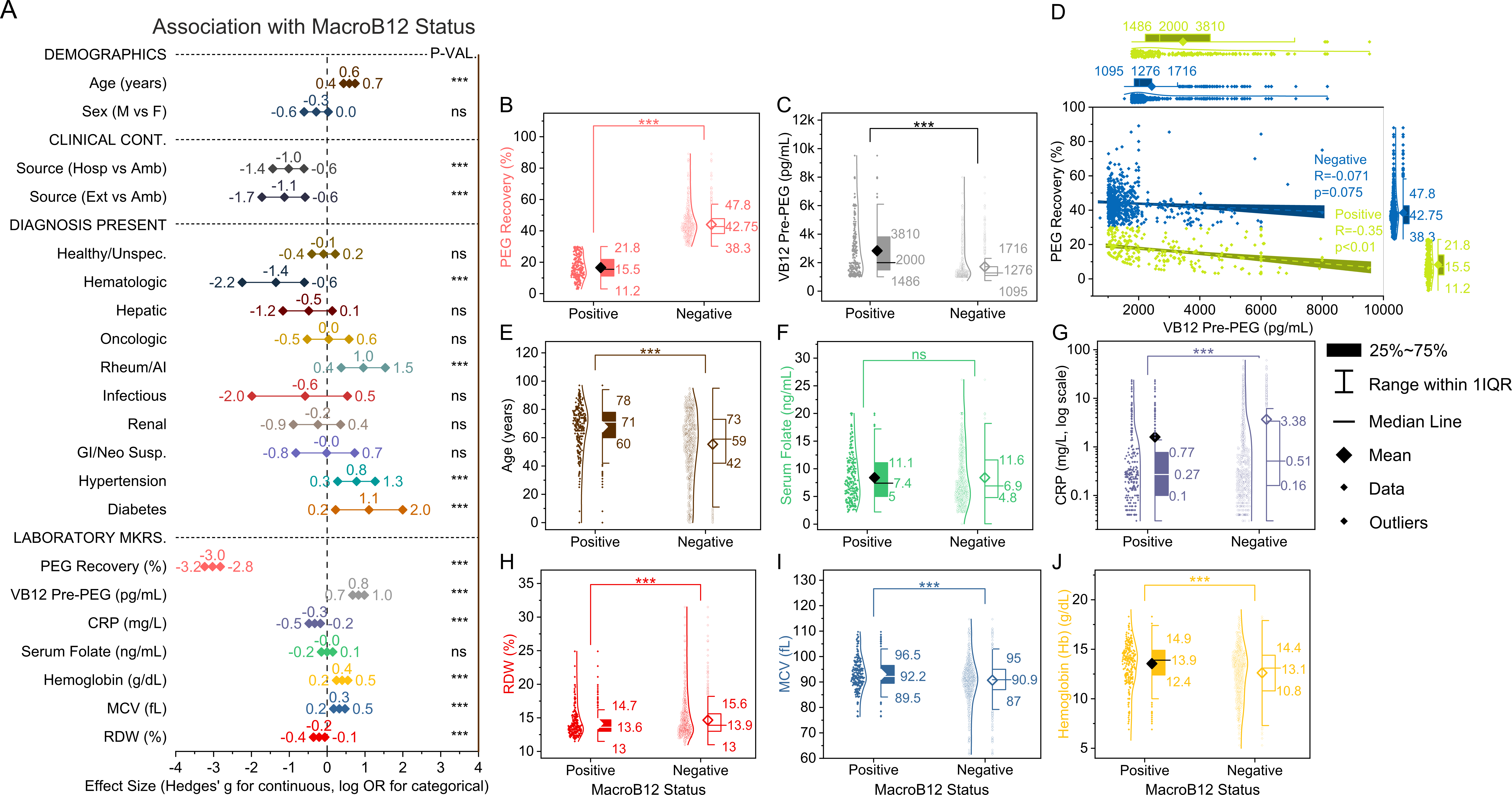}
      \caption{
        \textbf{Distinct biochemical and hematological signatures of MacroB12 status reveal paradoxical associations.}
        Comprehensive univariate analysis comparing MacroB12 negative (n=638) and positive (n=237) individuals. Box-and-violin plots show the full data distribution with median (horizontal line), mean (diamond), and interquartile range (box).   (\textbf{A}) A forest plot displays effect sizes (Hedges' g for continuous variables, log Odds Ratio for categorical) for all univariate associations.
        (\textbf{B}) PEG recovery percentages are markedly reduced in MacroB12 positive individuals, defining the analytical interference.
        (\textbf{C}) Pre-PEG Vitamin B12 concentrations are substantially elevated in the positive group.
        (\textbf{D}) A scatter plot illustrates the inverse correlation between pre-PEG B12 concentrations and PEG recovery, stratified by MacroB12 status.
        (\textbf{E}) The age distribution demonstrates a significant shift towards older individuals in the positive cohort.
        (\textbf{F}) Serum folate levels show no significant association, serving as an important negative control.
        (\textbf{G}) CRP levels are significantly lower in the positive group (log scale presentation).
        (\textbf{H}) RDW is modestly but significantly lower in the positive group.
        (\textbf{I}) MCV is paradoxically higher in MacroB12 positive individuals.
        (\textbf{J}) Hemoglobin levels are counterintuitively elevated in the positive cohort.  P-values from Wilcoxon rank-sum tests are displayed on relevant panels (***\textit{P} < 0.001; \textit{ns}, not significant).
    }

    \label{fig:boxviolins}  
\end{figure}

A comprehensive univariate analysis of the cohort delineates a complex landscape of factors associated with MacroB12 status. A forest plot summarising standardised effect sizes provides a panoramic quantitative overview (Fig.~\ref{fig:boxviolins}A), immediately establishing the dominant statistical signals. As anticipated, the largest effect sizes correspond to the definitional laboratory markers of MacroB12: a profoundly negative association with post-PEG recovery (Hedges’ g = -3.0; 95\% CI [-3.2-- -2.8]; \textit{P}<0.001) and a strongly positive association with pre-PEG Vitamin B12. A granular inspection of the underlying distributions confirms this profound analytical interference: MacroB12 positive cases present with pre-PEG B12 concentrations of 2000 pg/mL [1486--3810], substantially elevated compared to 1276 pg/mL [1095--1716] in the negative cohort (Fig.~\ref{fig:boxviolins}C). This directly results in markedly lower recovery percentages (Fig.~\ref{fig:boxviolins}B), with the inverse correlation between these two parameters being more pronounced in the positive group (Fig.~\ref{fig:boxviolins}D).

Far more compelling from a biological standpoint is the emergence of a paradoxical haematological signature that directly challenges the canonical presentation of functional Vitamin B12 deficiency. Rather than exhibiting signs of anaemia, MacroB12 positive individuals display significantly \textit{higher} Haemoglobin levels of 13.9 g/dL [12.4--14.9] (Fig.~\ref{fig:boxviolins}J). This is accompanied by an increased Mean Corpuscular Volume of 92.2 fL [89.5--96.5] (Fig.~\ref{fig:boxviolins}I), and a modestly but significantly \textit{lower} Red Cell Distribution Width of 13.6\% [13.0--14.7] (Fig.~\ref{fig:boxviolins}H). This counterintuitive profile suggests a more intricate underlying pathophysiology. It may represent a snapshot of a masked, sub-clinical deficiency yet to cause haematological decompensation, or, perhaps more plausibly, it could reflect a distinct patient sub-population with well-compensated systemic conditions that independently influence haematopoiesis, a hypothesis that finds convergent support from other clinical parameters.

Further analysis of the clinical context reinforces this latter hypothesis. The demographic data confirm a strong link to patient age, with a median of 78 years [71--82] in the positive group (Fig.~\ref{fig:boxviolins}E). Critically, MacroB12 status is strongly associated with a formal diagnosis in the Rheumatologic/Autoimmune category (log OR = 1.0; 95\% CI [0.4--1.5]; \textit{P}<0.001), while concurrently showing a significant inverse association with C-reactive protein levels, which were lower in the positive group at 0.27 mg/L [IQR: 0.1--0.77] (Fig.~\ref{fig:boxviolins}G). These findings—an association with autoimmune diagnoses in a state of low systemic inflammation—offer a consistent possible explanation for the observed haematological paradox. The specificity of these associations is reinforced by the absence of any significant difference in serum folate levels (median 7.4 ng/mL [5.0--11.1] vs 6.9 ng/mL [4.8--11.6]), which serves as a crucial negative control (Fig.~\ref{fig:boxviolins}F).

\subsection{Independent Predictors of MacroB12 Status}
\label{sec:multivariable_analysis}

\begin{figure}[ht!]
    \centering
     \includegraphics[width=1\textwidth]{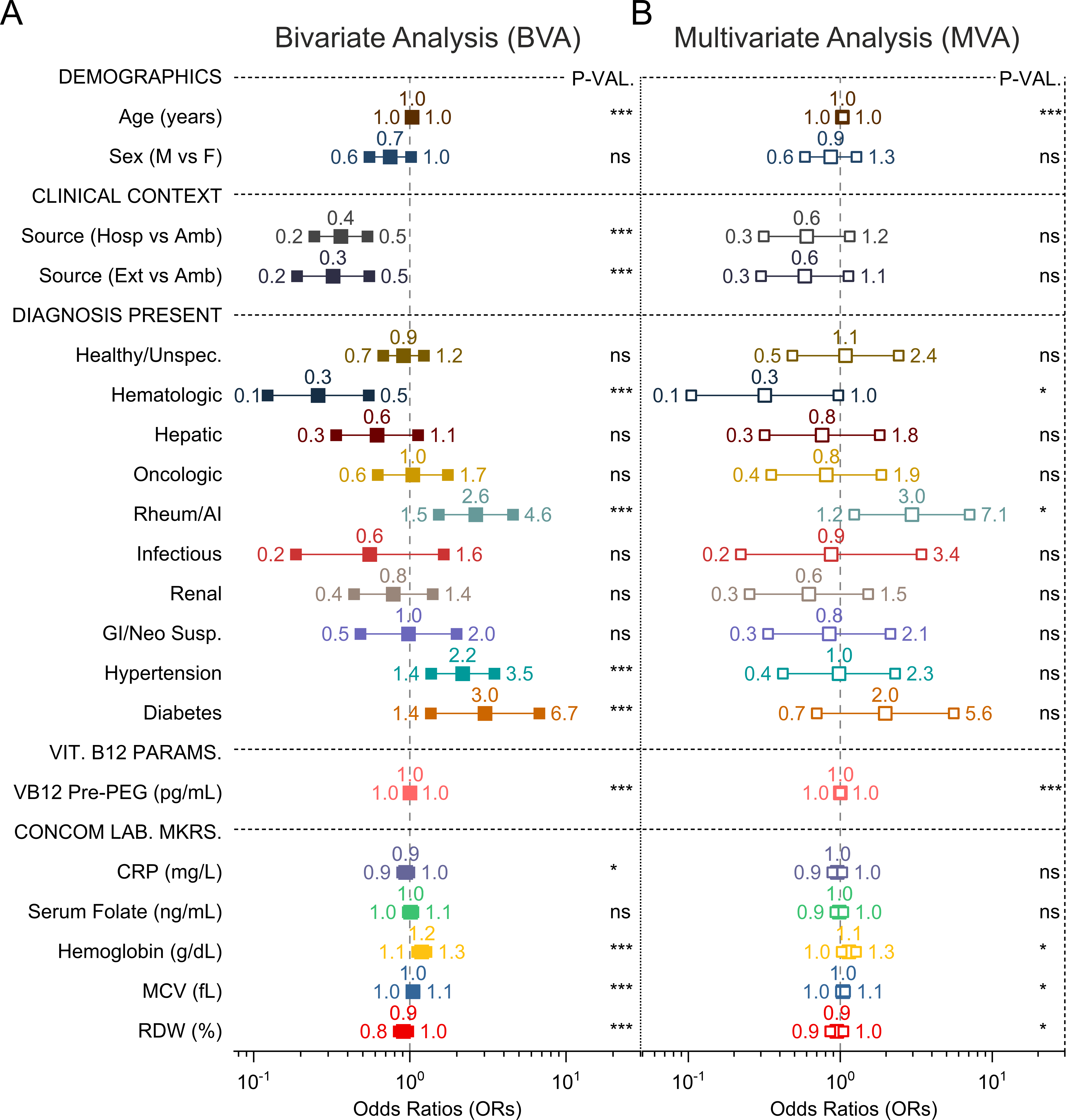}
    \caption{
        \textbf{Bivariate and Multivariate Forest Plots of Associations with MacroB12 Status.}
        The figure compares unadjusted and adjusted predictors of positive MacroB12 status, with Odds Ratios (ORs) presented on a logarithmic scale. (\textbf{A}) The Bivariate Analysis (BVA) panel displays unadjusted ORs for each variable, with solid squares representing the point estimate.      
        (\textbf{B}) The Multivariate Analysis (MVA) panel displays adjusted ORs derived from the final logistic regression model, which controlled for all variables shown. Hollow squares represent the point estimate.
        For both panels, horizontal lines indicate the 95\% confidence interval (CI), and the vertical dashed line represents the null effect (OR=1.0). Variables are grouped by clinical and laboratory category. P-values for each association are displayed on the right (\textsuperscript{***}P<0.001, \textsuperscript{*}P<0.05; ns, not significant).
    }
    \label{fig2:forest_plots} 
\end{figure}

To disentangle the complex correlations identified in the univariate analysis and isolate independent predictors of MacroB12 positivity, we first developed a multivariable logistic regression model (Fig.~\ref{fig2:forest_plots}). After adjusting for all potential confounders, several factors remained robustly associated. Older age was a significant predictor, with each additional year conferring a 3\% increase in the odds of having MacroB12 (adjusted Odds Ratio [aOR] 1.03; 95\% CI 1.02--1.04; \textit{P}<0.001). More notably, the paradoxical haematological signature persisted; both higher Haemoglobin (OR 1.14; 1.02--1.27) and higher Mean Corpuscular Volume (OR 1.04; 1.01--1.07) independently predicted increased odds. The clinical context was also critical: a formal Rheumatologic/Autoimmune diagnosis conferred a nearly threefold increase in odds (OR 2.96; 1.24--7.10), consistent with the hypothesis that MacroB12 often involves autoantibodies \cite{Bowen2006, Delgado2023}, whilst a Hematologic diagnosis was associated with significantly lower odds (OR 0.32; 0.10--0.97). Unsurprisingly, the pre-PEG B12 value remained the single most dominant predictor, with its effect being highly significant (\textit{P}<0.001).

\begin{figure}[ht!]
    \centering
     \includegraphics[width=1\textwidth]{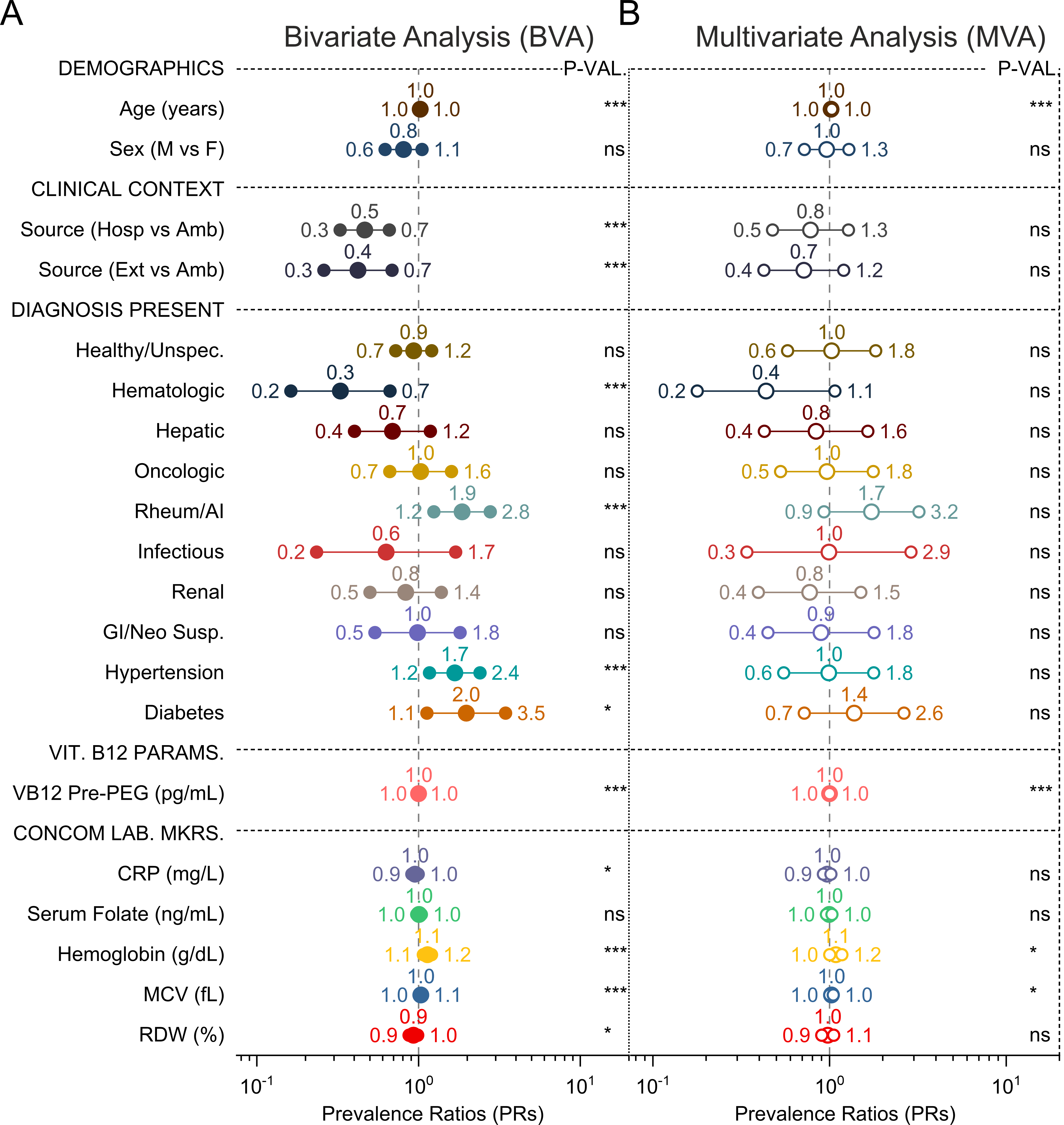}
    \caption{
        \textbf{Bivariate and Multivariate Analysis of Prevalence Ratios for MacroB12 Status.}
        The figure compares unadjusted and adjusted predictors of positive MacroB12 status, with Prevalence Ratios (PRs) presented on a logarithmic scale.
        (\textbf{A}) The Bivariate Analysis (BVA) panel displays the unadjusted PR for each variable, with solid circles representing the point estimate.
        (\textbf{B}) The Multivariate Analysis (MVA) panel displays the adjusted PR derived from the final multivariable Poisson regression model, which controlled for all variables shown. Hollow circles represent the point estimate.
        For both panels, horizontal lines indicate the 95\% confidence interval (CI), and the vertical dashed line represents the null effect (PR=1.0). Variables are grouped by clinical and laboratory category. P-values for each association are displayed on the right (\textsuperscript{***}P<0.001, \textsuperscript{*}P<0.05; ns, not significant).
    }
    \label{fig3:forest_plots_pr} 
\end{figure}

To provide a complementary perspective and generate a more conservative measure of effect, we developed a multivariable Poisson regression model to estimate Prevalence Ratios (PRs) (Fig.~\ref{fig3:forest_plots_pr}). This analysis yielded broadly consistent findings for the core predictors. Older age remained a strong predictor of higher prevalence (adjusted Prevalence Ratio [aPR] 1.02; 95\% CI 1.01--1.03; \textit{P}<0.001). Similarly, the key haematological markers, Haemoglobin (PR 1.09; 1.01--1.18) and MCV (PR 1.03; 1.00--1.05), retained their status as significant independent predictors of a higher prevalence of MacroB12. The overwhelming predictive power of the pre-PEG B12 concentration was, as expected, reaffirmed in this model as well.

A comparative assessment of the two models reveals a high degree of concordance for the principal continuous predictors, strengthening the evidence for their genuine association with MacroB12. However, a notable divergence emerges in the clinical diagnostic categories. In the Poisson model, the statistical significance for both the Rheumatologic/Autoimmune and Hematologic diagnoses was attenuated (\textit{P}>0.05 for both). This attenuation is an expected mathematical property when moving from ORs to PRs for outcomes with a high prevalence, as the OR can numerically overestimate the strength of an association when an outcome is not rare (i.e., prevalence >10\%), whereas the PR provides a more conservative and direct estimate of the prevalence increase \cite{Cummings2009, Zhang1998RR, McNutt2003}. Importantly, across both models, many factors significant in the univariate analysis—including patient source, sex, CRP, and RDW—were no longer significant after multivariable adjustment, highlighting their role as correlated observations rather than independent drivers.

\subsection{Diagnostic Performance of Pre-PEG Vitamin B12}
\label{sec:roc_analysis}

\begin{figure}[ht!]
    \centering
    \includegraphics[width=\textwidth]{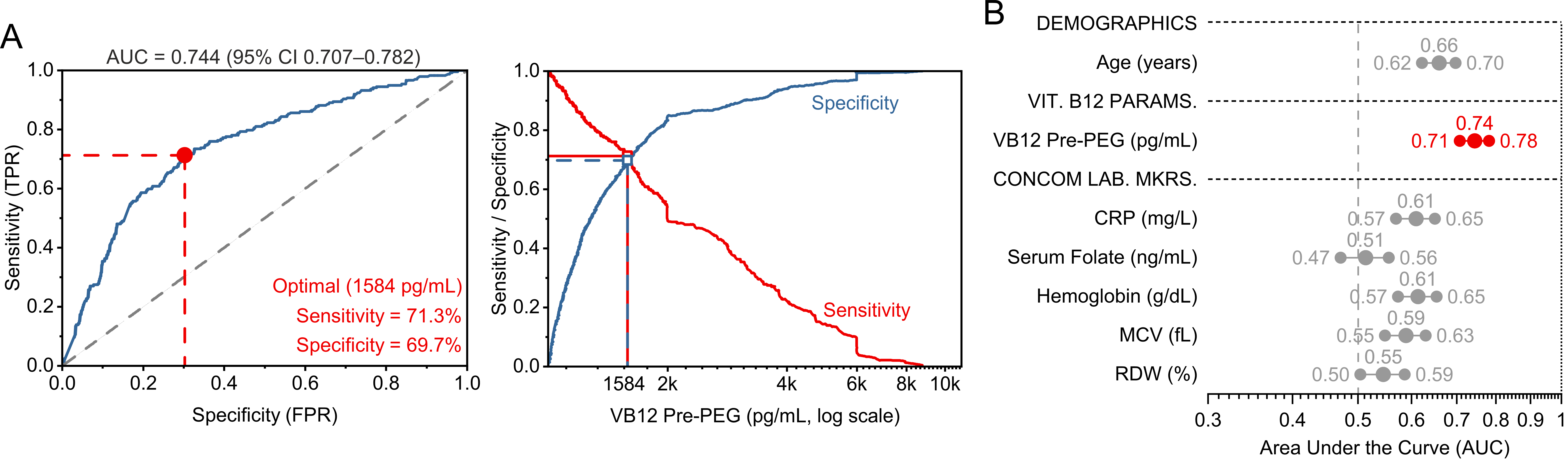}
    \caption{
        \textbf{Diagnostic Performance of Pre-PEG Vitamin B12 and Comparison with Other Markers.}
        (\textbf{A}) Receiver Operating Characteristic (ROC) analysis for pre-PEG B12 concentration. The left plot shows the ROC curve, yielding an Area Under the Curve (AUC) of 0.744 (95\% CI 0.707--0.782). The red point indicates the optimal threshold (1584 pg/mL) identified via the Youden index, with a corresponding sensitivity of 71.3\% and specificity of 69.7\%. The plot on the right displays sensitivity and specificity across the full range of thresholds, with the optimal cut-point marked by a vertical dashed line.
        (\textbf{B}) Forest plot comparing the AUCs for pre-PEG B12 and other potential predictors. Pre-PEG B12 (highlighted in red) demonstrates superior discriminatory power compared to the other demographic and laboratory markers.
    }
    \label{fig:roc_main} 
\end{figure}

To quantify the clinical utility of pre-PEG Vitamin B12 as a standalone diagnostic marker for MacroB12 status, we conducted a Receiver Operating Characteristic (ROC) analysis (Fig.~\ref{fig:roc_main}). The analysis revealed a moderate but clinically significant predictive utility, with an Area Under the Curve (AUC) of 0.744 (95\% CI 0.707--0.782), indicating a performance substantially better than chance (Fig.~\ref{fig:roc_main}A). Application of the Youden index identified an optimal diagnostic threshold of 1584 pg/mL. At this threshold, the test provides a well-balanced performance, yielding a sensitivity of 71.3\% and a specificity of 69.7\%. The trade-off between sensitivity and specificity across all possible thresholds is further visualised in the adjacent plot, confirming the selection of this optimal cut-point.

To place this performance in a practical context, we compared the discriminatory power of pre-PEG B12 against that of other readily available demographic and laboratory parameters (Fig.~\ref{fig:roc_main}B). This comparison unequivocally establishes pre-PEG B12 as the most powerful single predictor. Its AUC of 0.744 markedly exceeds that of the next best individual predictor, patient age (AUC 0.66; 95\% CI 0.62--0.70). Other haematological and biochemical markers, such as Haemoglobin (AUC 0.61; 0.57--0.65) and CRP (AUC 0.61; 0.57--0.65), offered only modest utility, while serum folate provided negligible predictive value (AUC 0.51; 0.47--0.56), approaching the line of chance. These results validate the central role of pre-PEG B12 measurement in any diagnostic algorithm for MacroB12, while the moderate absolute performance highlights the potential for improved accuracy through the integration of multiple factors in a comprehensive predictive model. Further analyses examining factors correlated with the PEG recovery percentage metric itself and exploring variable importance using machine learning methods are presented in the Supplementary Information (Supplementary Fig.~S2).

\section{Discussion}\label{sec:discussion}

MacroB12 represents a well-recognized yet often challenging analytical confounder in Vitamin B12 diagnostics. Its presence, typically involving B12 complexed with immunoglobulins, can artifactually elevate measured serum levels, obscuring potential deficiency states and complicating clinical interpretation \cite{Bowen2006, Delgado2023}. This analytical challenge is superimposed upon an existing landscape where even the precise population frequency and full clinical spectrum of true Vitamin B12 deficiency remain subjects of considerable deliberation, a point underscored by observations of a ``shared ignorance'' regarding these fundamental aspects \cite{Naylor2002Letter}. While polyethylene glycol (PEG) precipitation remains a pragmatic method for detecting MacroB12, its widespread implementation is not universal, motivating the search for readily available clinical or laboratory parameters that might predict MacroB12 positivity. Leveraging a substantial cohort (N=875) with multiply imputed data to ensure robustness against missingness, this study systematically evaluated a comprehensive set of potential correlates associated with MacroB12 status, defined by the standard <30\% PEG recovery criterion.

\begin{figure}[htbp!]
    \centering
    \includegraphics[width=\textwidth]{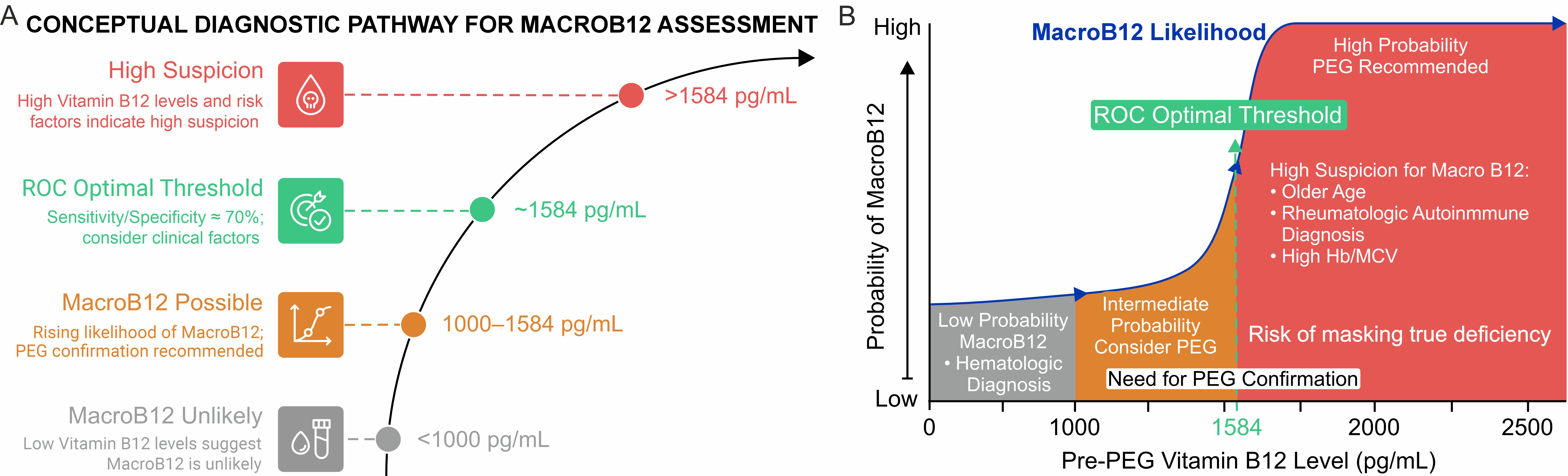}
    \caption{\textbf{A Proposed Conceptual Diagnostic Pathway} and Risk Stratification for MacroB12 Assessment.
    \textbf{(A)} A conceptual diagnostic pathway illustrating escalating suspicion for MacroB12 based on pre-PEG Vitamin B12 levels. Key thresholds include <1000 pg/mL (MacroB12 unlikely), 1000--1584 pg/mL (MacroB12 possible, PEG confirmation recommended), the ROC optimal threshold of ~1584 pg/mL (Sensitivity/Specificity $\approx$ 70\%), and >1584 pg/mL (high suspicion). Icons represent clinical decision points or states.
    \textbf{(B)} Schematic representation of the estimated probability of MacroB12 increasing with pre-PEG Vitamin B12 levels. Colored zones indicate low (<1000 pg/mL, grey), intermediate (1000--1584 pg/mL, orange), and high (>1584 pg/mL, red) probability/suspicion strata. The ROC optimal threshold (1584 pg/mL) is marked with a dashed line. Key clinical correlates identified from multivariable analysis that modulate suspicion within the high probability zone (e.g., older age, rheumatologic/autoimmune diagnosis, high Hb/MCV) and a factor associated with lower likelihood (hematologic diagnosis) are annotated. PEG confirmation is recommended for intermediate and high probability strata, especially considering the risk of masking true B12 deficiency at very high measured levels.}
    \label{fig:diagnostic_pathway}
\end{figure}

Unsurprisingly, the pre-PEG Vitamin B12 concentration emerged as the dominant factor associated with eventual MacroB12 classification. While tautological to some extent due to the nature of the interference, our multivariable regression models rigorously confirmed this association persists independently of numerous potential confounders (Fig.~\ref{fig2:forest_plots}). From a clinical utility standpoint, however, the discriminatory power of the pre-PEG level alone, although statistically significant, proved only moderate (AUC 0.744 (0.707--0.782); Fig.~\ref{fig:roc_main}A). The Youden-optimal 1584.0~pg/mL threshold offered a reasonable balance (Sensitivity 71.3\% and Specificity 69.7\%). Still, its performance indicates that relying solely on such a cutoff would lead to an unacceptable rate of misclassification in clinical practice~\cite{Nahm2016}. This finding strongly reinforces that while exceptionally high B12 levels warrant suspicion, confirmatory testing remains indispensable for definitive MacroB12 identification, particularly when clinical indices suggest deficiency. This graded risk and the interplay of factors are visually synthesized in Figure~\ref{fig:diagnostic_pathway}, which proposes a conceptual diagnostic pathway integrating pre-PEG B12 levels with key clinical correlates.

Beyond the intrinsic link with measured B12, our multivariable analysis identified two key patient profiles independently associated with MacroB12 status. The first and most mechanistically significant is a profile suggestive of autoimmunity. Both increased age (multivariable-adjusted aOR~1.03 (1.02--1.04)) and a formal Rheumatologic/Autoimmune diagnosis (aOR~2.96~(1.24--7.10)) were strong predictors (Fig.~\ref{fig2:forest_plots}B). This robustly supports the prevailing hypothesis that MacroB12 is frequently an antibody-mediated phenomenon~\cite{Bowen2006, Delgado2023, Herrmann2003} and has direct clinical implications: it defines a high-risk cohort for targeted screening and suggests MacroB12 could serve as a biomarker for underlying autoimmune processes.

The second profile presents a compelling hematological paradox. MacroB12 positivity was independently associated with \textit{higher} hemoglobin (aOR~1.14 (1.02--1.27)) and \textit{higher} Mean Corpuscular Volume (aOR~1.04 (1.01--1.07)), contradicting the classic presentation of B12 deficiency. Several non-mutually exclusive mechanisms may explain this counterintuitive signature: (i) masking of an early-stage functional deficiency before hematological decompensation; (ii) distinct baseline hematological parameters in patients with autoimmune conditions; (iii) selection bias in our referral cohort, enriching for patients investigated for reasons other than anemia; or (iv) a correlation with robust immunoglobulin production and better overall nutritional status. \textbf{Disentangling these possibilities represents a critical direction for future research.} Prospective studies are urgently needed, incorporating direct functional markers of B12 status (e.g., methylmalonic acid, homocysteine) alongside comprehensive autoantibody panels. Such studies would allow for stratification of this paradoxical association by clinical context, such as the presence or absence of specific autoimmune diseases, to elucidate the underlying pathophysiology.

These findings, integrated into our proposed risk stratification model (Fig.~\ref{fig:diagnostic_pathway}B), highlight the importance of clinical context. The reduced odds of MacroB12 in patients with a Hematologic diagnosis (aOR~0.32~(0.10--0.97)) likely reflects an ascertainment bias towards true deficiency in that group. Furthermore, the necessity of multivariable adjustment was underscored by the attenuation of strong bivariate associations for factors like patient source and CRP (Fig.~\ref{fig:boxviolins}, Fig.~\ref{fig2:forest_plots}A vs.~\ref{fig2:forest_plots}B), confirming their role as correlates rather than independent drivers.

Further context was provided by exploratory analyses detailed in the supplement. Beta regression modelling of the quantitative PEG recovery metric itself identified not only the expected inverse relationships with pre-PEG B12 and age but also positive associations with CRP and certain diagnostic groups (Oncologic, Healthy/Unspecified) (Supplementary Fig.~S2A, B). The biological underpinnings of these latter associations remain speculative and could relate to non-specific protein interactions influencing precipitation efficiency. Complementary Random Forest analysis confirmed the paramount predictive importance of pre-PEG B12, age, and hemoglobin for MacroB12 classification (Supplementary Fig.~S2C), reinforcing the key variables identified via regression while capturing potential non-linear contributions.

Recognizing the variability in PEG recovery thresholds reported in the literature for MacroB12 identification \cite{Delgado2024, Remacha2014}, we performed a sensitivity analysis employing an alternative cutoff of <40\% (Supplementary Table~S3). This alternative definition resulted in an increased prevalence of MacroB12 positivity (50.9\% vs. 27.1\% with the <30\% threshold) and, while core associations for pre-PEG B12, age, and Rheumatologic/Autoimmune diagnoses remained robustly significant, the statistical significance for hemoglobin, MCV, and Hematologic diagnosis was attenuated. Notably, Diabetes Mellitus emerged as a significant predictor under this broader definition (aOR 4.53, 95\% CI 1.32--15.5). The discriminatory performance of pre-PEG B12, when assessed against this <40\% outcome definition, yielded an AUC of 0.713 (95\% CI 0.679--0.746), with an optimal threshold of 1579.5 pg/mL (Sensitivity 58.2\%, Specificity 75.3\%). These findings suggest that while our primary conclusions regarding the strongest predictors are stable, the precise effect estimates and the significance profile of secondary factors can be influenced by the chosen diagnostic threshold, underscoring the importance of standardized definitions or context-specific threshold validation.

The contributions of this study are supported by several key methodological strengths that ensure both statistical robustness and clinical relevance. Most notably is the large cohort size (N=875), which, to our knowledge, is one of the largest populations in the literature for this particular analytical question. This provided the statistical power needed to go beyond simple univariate associations and develop comprehensive multivariable models that can identify independent predictors from a complex array of clinical variables. Secondly, we directly addressed the challenge of real-world evidence by using Multiple Imputation by Chained Equations. This method preserved the integrity of our entire cohort by avoiding the selection bias and loss of power caused by listwise deletion of cases with missing data. Finally, our multi-faceted analytical framework was carefully designed to suit the data structure: we used logistic regression for our primary binary outcome, supplemented it with modified Poisson regression to offer more conservative and interpretable prevalence ratios, and applied Beta regression to appropriately model the challenging proportional nature of the PEG recovery data. Overall, this rigorous approach sets a strong methodological example for future research in the field.

Nevertheless, certain limitations must be considered as they frame the necessary directions for future investigations. Our primary limitation is the single-center design, which necessitates caution in generalizing our findings. While our cohort is large, it was derived from a single tertiary referral center in a specific geographic region (Canary Islands, Spain), which may have unique population demographics, genetic backgrounds, and clinical referral patterns. Therefore, the precise effect estimates (e.g., adjusted Odds Ratios) and, most importantly, the specific optimal threshold for pre-PEG B12 of 1584.0~pg/mL, should be interpreted as a robust, data-derived starting point rather than a universally applicable cutoff. We strongly recommend that this threshold and our broader predictive findings undergo external validation. Further limitations include our reliance on administrative diagnostic codes, which inherently limits clinical granularity, and the fact that the MacroB12 definition was predicated solely on the PEG recovery assay rather than a biochemical reference standard like gel filtration chromatography. Finally, the cross-sectional nature of the data precludes any inference regarding causality or the temporal evolution of MacroB12.

\section{Conclusions}\label{sec:conclusions}

This investigation delineated the clinical and laboratory landscape associated with MacroB12, a frequent source of analytical interference in Vitamin B12 assessment, within a large, well-characterized cohort utilizing multiply imputed data. Through comprehensive multivariable regression and ROC analyses, we established the significance of pre-precipitation B12 levels, patient age, specific hematological indices (hemoglobin, MCV), and relevant clinical contexts, particularly Rheumatologic/Autoimmune diagnoses, as key factors independently correlated with MacroB12 positivity defined by PEG precipitation. While quantifying the moderate discriminatory power of pre-PEG B12 levels (AUC~0.74), our work also underscores the limitations of relying solely on this marker for definitive classification.

The findings confirm elevated pre-PEG Vitamin B12 levels and older age as robust, independent correlates of MacroB12 positivity. The strong association identified with Rheumatologic/Autoimmune diagnoses lends considerable support to an underlying autoimmune etiology, likely antibody-mediated, in a significant proportion of individuals presenting with this phenomenon. Although pre-PEG B12 concentration offers moderate discriminatory utility, its performance characteristics (sensitivity ~71\%, specificity ~70\% at the optimal threshold) highlight the continuing necessity for specific confirmatory assays, primarily PEG precipitation, to reliably diagnose MacroB12 and avoid clinical misinterpretation, especially in the presence of associated factors like advanced age or autoimmune conditions. Illuminating the precise mechanisms connecting MacroB12 with the observed alterations in hemoglobin and MCV, alongside factors influencing quantitative PEG recovery, remain important avenues for future research. The study underscores the need for prospective studies incorporating detailed clinical phenotyping and direct autoantibody measurements. Ultimately, the development of integrated prediction models that leverage multiple readily available parameters holds promise for optimizing clinical workflows, improving the identification of patients requiring confirmatory MacroB12 testing, and enhancing the overall diagnostic accuracy for assessing actual Vitamin B12 status in the face of this common and often perplexing analytical challenge.

\section{Materials and Methods}\label{sec:methods}
\subsection{Study Design, Population, and Ethical Considerations}
This investigation employed a retrospective, observational cohort design to comprehensively evaluate clinical and laboratory parameters associated with MacroB12 status. The sequential process guiding patient identification, eligibility screening, data acquisition, processing, and subsequent analytical strategies is detailed schematically in Figure~\ref{fig:study_flow}. The source population encompassed all individuals (N = 115,460) for whom a serum Vitamin B12 (B12) measurement was clinically requested between 1\textsuperscript{st} June 2023 and 31\textsuperscript{st} May 2024, across diverse clinical settings including primary care, specialized outpatient consultations, and inpatient services at the University Hospital Nuestra Señora de Candelaria, a tertiary referral center in Santa Cruz de Tenerife, Spain (Fig.~\ref{fig:study_flow}, Step 1). From this initial pool, we applied inclusion criteria, specifically selecting individuals with serum B12 levels exceeding \SI{1000}{pg/mL} (N = 2,059), a threshold often employed to trigger further investigation for potential interferences like MacroB12 (Fig.~\ref{fig:study_flow}, Step 2). To ensure the measured B12 levels reflected endogenous status, patients with documented administration of B12 supplements recorded in their clinical history were subsequently excluded (N excluded = 1,073; N remaining = 986) (Fig.~\ref{fig:study_flow}, Step 3). Following these selection steps and the further exclusion of 111 individuals due to missing outcome data (MacroB12 status derived from PEG recovery, necessary for subsequent modeling), the final analytical cohort for imputation and primary analyses comprised 875 unique individuals (Fig.~\ref{fig:study_flow}, Steps 4 and 5).

Demographic variables (age at time of B12 test, sex) and extensive clinical information were meticulously retrieved for the selected cohort from the institutional electronic health record platforms, DragoAE (hospital) and DragoAP (primary care) (Fig.~\ref{fig:study_flow}, Step 6). Data extraction prioritized documented clinical diagnoses potentially associated with altered B12 metabolism or the presence of interfering substances, systematically screening for codes related to hematologic neoplasms (e.g., myeloproliferative disorders, lymphoma, leukemia), solid tumors, significant hepatic disease (e.g., cirrhosis, hepatitis), chronic kidney disease, and autoimmune or rheumatologic conditions (e.g., rheumatoid arthritis, systemic lupus erythematosus, inflammatory bowel disease). Recognizing regional health priorities and potential confounders, specific searches were also conducted for diagnoses of diabetes mellitus and recorded clinical suspicion of underlying neoplasm (e.g., investigation for unexplained weight loss). Concomitant laboratory results proximate to the index B12 measurement, including serum folate, complete blood count parameters (hemoglobin [Hb], mean corpuscular volume [MCV], red cell distribution width [RDW]), and C-reactive protein (CRP), were extracted from the interfacing OpenLab Laboratory Information System (Fig.~\ref{fig:study_flow}, Step 6). Standard data curation practices involved rigorous de-duplication based on unique patient identifiers (medical record number) and test dates, followed by complete anonymization of all patient-level data prior to statistical analysis, in strict adherence to institutional data protection protocols (Fig.~\ref{fig:study_flow}, Step 7).

This study was conducted in full accordance with the ethical principles for medical research involving human subjects as outlined in the Declaration of Helsinki and its subsequent amendments. The study protocol, titled ``Estudio sobre la presencia de macrocomplejos de vitamina B12 en nuestro ámbito hospitalario'' (version 2, 12/02/2024), under protocol code CHUNSC\_2023\_127, was formally evaluated and received a favorable ethical opinion from the \textit{Comité de Ética de la Investigación con medicamentos del Complejo Hospitalario Universitario de Canarias (Provincia de Santa Cruz de Tenerife), Spain} (Ethics Committee for Research with Medicines of the University Hospital Complex of Canarias, Santa Cruz de Tenerife Province, Spain). The approval was granted during the committee's session on January 25, 2024.

The Ethics Committee confirmed that the protocol met the necessary requirements for suitability in relation to the study's objectives and affirmed that the investigator's capacity and available resources were adequate for conducting the research without infringing upon ethical postulates. Given the retrospective observational design of the study, which relied exclusively on the analysis of routinely collected clinical and laboratory data, and the assurance of personal data confidentiality, the Ethics Committee exceptionally determined that individual informed consent would not be solicited from participants. Robust measures were implemented throughout the study to ensure patient data confidentiality and to comply with all applicable institutional and national data protection regulations, including the Spanish Organic Law on Data Protection and Guarantee of Digital Rights (LOPDGDD 3/2018), as stipulated in the ethical approval. All research activities were conducted in compliance with the approved protocol and relevant ethical guidelines.

\subsection{Biochemical Analyses and MacroB12 Assessment}
Quantitative determination of total serum B12 concentrations was performed within the centralized Clinical Analysis Laboratory of the University Hospital Nuestra Señora de Candelaria. The laboratory employs an automated chemiluminescent microparticle immunoassay (CMIA) utilizing purified intrinsic factor for B12 capture, executed on the Alinity i analyzer platform (Abbott Diagnostics, Abbott Park, IL, USA), according to standardized operating procedures. For samples yielding results exceeding the instrument's upper limit of quantification, automated serial dilutions recommended by the manufacturer were performed by the system until a precise B12 concentration within the validated linear measurement range was obtained. Adherence to pre-analytical guidelines was maintained; samples were processed on the day of blood collection whenever feasible. If immediate analysis was precluded, serum samples were promptly refrigerated at \SI{4}{\celsius} and analyzed within the manufacturer-specified stability window.

To specifically investigate the presence of MacroB12 complexes, which interfere with standard immunoassays, polyethylene glycol (PEG) precipitation was systematically performed on all serum samples meeting the initial inclusion criteria (pre-treatment B12 > \SI{1000}{pg/mL}). The procedure involved mixing \SI{200}{\micro\liter} of patient serum with an equal volume (\SI{200}{\micro\liter}) of a 25\% (w/v) PEG 6000 (Sigma-Aldrich, St. Louis, MO, USA) solution prepared in phosphate-buffered saline. The serum-PEG mixture was thoroughly vortexed to ensure adequate mixing and subsequently centrifuged at high speed (\SI{10800}{rpm}, specific g-force dependent on rotor radius) for 5 minutes to effectively precipitate large molecular weight complexes, including immunoglobulins potentially bound to B12. The resulting clear supernatant, theoretically enriched in free or normally bound (transcobalamin) B12 and depleted of MacroB12 complexes, was carefully collected. The B12 concentration within this supernatant was then re-measured using the identical Alinity i CMIA methodology.

The efficiency of immunoglobulin-complex precipitation and the relative amount of non-precipitated B12 were quantified by calculating the percentage recovery using the standard formula, accounting for the 1:1 dilution introduced by PEG addition:
\[
    \text{Recovery (\%)} = \frac{\text{B12 concentration in supernatant after PEG} \times 2}{\text{B12 concentration in serum before PEG}} \times 100
\]
Patient samples were then definitively classified based on this recovery percentage for the primary study outcome. Informed by established literature precedent suggesting PEG recovery thresholds between <30\% and <40\% for MacroB12 identification \cite{Delgado2024, Remacha2014, Soleimani2019}, and further supported by internal laboratory validation data derived from analogous precipitation assays (e.g., macroprolactin), samples exhibiting <30\% recovery were defined as MacroB12 positive, indicating significant immunoassay interference presumably due to large B12-immunoglobulin complexes. Conversely, samples with >60\% recovery were classified as MacroB12 negative, suggesting minimal interference. The intermediate recovery range of 30--60\% was deemed inconclusive for definitive binary classification and these patients were excluded from analyses requiring dichotomous MacroB12 status, although their quantitative recovery data were retained for relevant continuous analyses. This operational definition, utilizing the <30\% threshold, resulted in the final analytical cohort of 875 individuals after excluding those with missing recovery data or inconclusive results (Fig.~\ref{fig:study_flow}, Step 5).

To ensure the reliability and validity of the PEG precipitation procedure throughout the study period, an internal quality control (QC) measure was implemented. A pooled serum sample, created from residual specimens of patients known to be receiving high-dose B12 supplementation (expected to contain predominantly free or normally bound B12), was aliquoted. With each analytical batch involving patient sample PEG precipitation, an aliquot of this internal QC pool underwent identical PEG treatment and B12 measurement. Across all runs, the post-PEG B12 levels in the QC material consistently fell within the expected physiological reference range, and recovery percentages were uniformly >60\%, providing continuous validation of the precipitation process's effectiveness in removing non-complexed B12 and confirming assay integrity.

\subsection{Statistical Analysis and Visualization}\label{sec:stats}

All statistical analyses were conducted using R software (version 4.3.2; R Core Team, Vienna, Austria), leveraging functionalities from established packages including \texttt{mice} (v3.16.0) for imputation, \texttt{betareg} (v3.1-5) for Beta regression, \texttt{ranger} (v0.16.0) and \texttt{vip} (v0.4.1) for Random Forest analysis, \texttt{pROC} (v1.18.5) for ROC analysis, and the \texttt{tidyverse} suite (v2.0.0) for general data manipulation and visualization \cite{RCoreTeam, vanBuuren2011mice, CribariNeto2004, Wright2017}. Statistical significance was defined a priori as a two-sided P-value < 0.05 for all inferential tests.

\begin{figure}[ht!]
    \centering
    \includegraphics[width=1\textwidth]{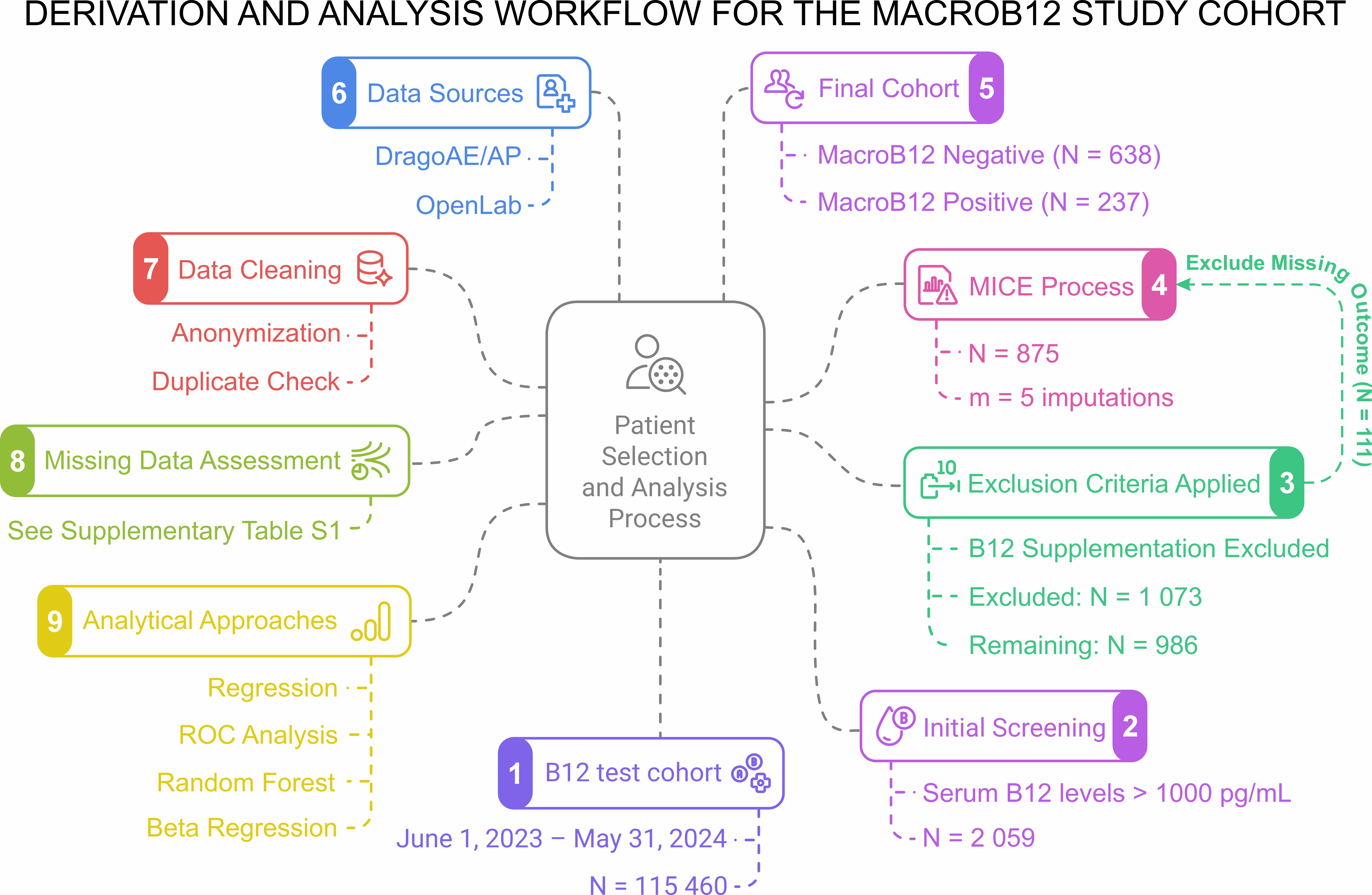} 
    \caption{\textbf{Derivation and Analysis Workflow for the MacroB12 Study Cohort.} Schematic overview illustrating the identification of the initial patient pool from B12 test requests (June 2023-May 2024), application of inclusion (>1000 pg/mL) and exclusion (B12 supplementation) criteria, data sourcing and processing steps including multiple imputation (MICE), derivation of the final analytical cohort (N=875) stratified by MacroB12 status (<30\% PEG recovery), and the subsequent statistical analysis approaches employed. Numbers indicate sample size at specific stages where determined (final N=875 entering MICE). See Supplementary Table~S1 for details on missing data assessment.}
    \label{fig:study_flow}
\end{figure}

Given the retrospective nature of data collection from routine clinical practice, missing data were anticipated and formally assessed (Supplementary Table~S1). Notable levels of missingness were confirmed for C-reactive protein (CRP, 49.7\%) and serum folate (20.6\%), with lower but non-negligible missingness for hemoglobin, MCV, and RDW (all 7.2\%). To address potential bias and loss of statistical power associated with listwise deletion, we employed Multiple Imputation by Chained Equations (MICE) \cite{vanBuuren2011mice}. The imputation model incorporated all predictor variables intended for inclusion in multivariable models (listed in Table~\ref{tab:baseline_char}) as well as the primary outcome variable (MacroB12 Status) and the quantitative PEG Recovery (\%), adhering to best practices recommending inclusion of the outcome. Five complete datasets ($m=5$) were generated using the \texttt{mice} package (v3.16.0). Predictive mean matching (PMM, k=5 nearest neighbors) served as the imputation method for continuous variables, logistic regression (\texttt{logreg}) for binary variables (sex, diagnostic flags), and polytomous logistic regression (\texttt{polyreg}) for the three-level patient source variable. Convergence was assessed after 10 iterations via visual inspection of trace plots, confirming stability of the imputed chains. All subsequent statistical analyses involving potentially missing predictors were performed independently on each of the five imputed datasets, and the resulting parameter estimates (e.g., regression coefficients, odds ratios) and their variances were combined using Rubin’s rules to obtain final pooled estimates, confidence intervals, and P-values \cite{Rubin1987} (Fig.~\ref{fig:study_flow}, Step 4).

Baseline characteristics were compared between MacroB12 positive and negative groups using Wilcoxon rank-sum tests for continuous variables (reported as Median [IQR]) and Chi-squared or Fisher’s exact tests for categorical variables (reported as N (\%)). To identify independent predictors of MacroB12 positivity (binary outcome, prevalence 27.1\%), pooled multivariable logistic regression analyses were conducted, yielding adjusted Odds Ratios (aOR) with 95\% CIs. Considering the relatively high prevalence, pooled modified Poisson regression models with robust variance estimation were also fitted to directly estimate adjusted Prevalence Ratios (aPR) with 95\% CIs \cite{Zou2004modified}. Factors associated with the quantitative PEG Recovery (\%), conceptualized as a proportion bounded between 0 and 1, were evaluated using pooled multivariable Beta regression with a logit link, applied to recovery data appropriately transformed to the (0,1) interval \cite{Smithson2006} (see Supplementary Fig.~S2A, B for results and diagnostics).

The discriminatory ability of key continuous variables for MacroB12 status was quantified using Receiver Operating Characteristic (ROC) curve analysis via the \texttt{pROC} package, performed on the first imputed dataset for visualization and threshold determination. The Area Under the Curve (AUC) with bootstrapped 95\% CIs served as the primary performance metric. Optimal classification thresholds balancing sensitivity and specificity were identified using the Youden index (J = Sensitivity + Specificity - 1). Finally, to complement association analyses and explore potentially non-linear contributions, the relative importance of all predictors in classifying MacroB12 status was assessed using Random Forest models (ranger implementation, $n_{trees}=500$), averaging permutation-based importance scores across the five imputed datasets via the \texttt{vip} package (Supplementary Fig.~2C).

\textit{Sample Size Considerations:} To contextualize the study's ability to detect effects of clinical relevance, post-hoc estimations of the Minimum Detectable Effect Size (MEDS) were performed for key variable types, setting statistical power at 80\% and \(\alpha=0.05\). Calculations based on the final cohort size (N=875, derived from the first imputation) indicated sensitivity to detect small standardized mean differences for continuous variables (Cohen's d \(\approx\) 0.21) and very small standardized effects for binary predictors within logistic regression frameworks (Cohen's f\textsuperscript{2} \(\approx\) 0.009). These estimates suggest the study was well-powered for moderate effects but may have been underpowered to reliably detect very subtle associations (see Supplementary Table~S2 for detailed MEDS).

The conceptual diagnostic pathway schematic (Figure~\ref{fig:diagnostic_pathway}) and the workflow diagram (Figure~\ref{fig:study_flow}) were prepared
using Inkscape (version 1.3.2; Inkscape Project, 2024~\cite{InkscapeProject2024}).

\section{Declarations}

\subsection*{Funding}
J.M. G.-N. has been supported by a predoctoral contract from the Centro de Investigación Biomédica en Red de Epidemiología y Salud Pública (CIBERESP, Instituto de Salud Carlos III). R.F. acknowledges partial support from the María Zambrano-Senior grant (Spanish Ministerio de Universidades and Next-Generation EU); Grant C-EXP-265-UGR23 funded by Consejería de Universidad, Investigación \& Innovación \& ERDF/EU Andalusia Program; Grant PID2022-137228OB-I00 funded by the Spanish Ministerio de Ciencia, Innovación y Universidades, MICIU/AEI/10.13039/501100011033 \& "ERDF/EU A way of making Europe"; and the Modelling Nature Research Unit, project QUAL21-011.

\subsection*{Conflict of Interest}
The authors declare that they have no financial or personal relationships that could be construed as potential conflicts of interest with respect to the research presented in this manuscript.

\subsection*{Ethics Approval and Informed Consent Statement}
The study protocol was officially reviewed and received a favourable ethical opinion from the Ethics Committee for Research with Medicines at the University Hospital Complex of Canarias (Protocol Code: CHUNSC\_2023\_127). The committee approved a waiver for the requirement of individual informed consent. This decision was based on the retrospective observational nature of the study, which solely relied on the analysis of fully anonymised, routinely collected clinical and laboratory data, ensuring that patient confidentiality was preserved throughout.

\subsection*{Data and Code Availability} %
To foster transparency and reproducibility, the primary individual participant dataset (N=875) is openly accessible. This dataset---specifically the first of five multiply imputed datasets used for generating representative descriptive statistics and key visualisations---has undergone a rigorous de-identification process to comprehensively protect patient confidentiality. It, along with the complete R scripts utilised for all stages of data cleaning, multiple imputation, statistical analysis, and figure generation, can be found in our GitHub repository: \href{https://github.com/renee29/Predicting-MacroB12-R-Analysis}{https://github.com/renee29/Predicting-MacroB12-R-Analysis}. This entire replication package---including a comprehensive data dictionary detailing all variables, their coding, and units---has also been permanently archived and is citable via Zenodo (DOI: \href{https://doi.org/10.5281/zenodo.15420222}{10.5281/zenodo.15420222}). The complete MICE imputation object (\texttt{mice\_imputation\_object.rds})---which formed the basis for all pooled multivariable analyses and ensures precise replication---can be made available by the corresponding author upon reasonable request, subject to appropriate data governance and ethical considerations. The original source clinical data are not publicly available due to inherent patient-identifying information and in accordance with ethical approvals designed to ensure privacy; however, access for specific collaborative research proposals---following institutional review board approval and data use agreements---may be considered.

\subsection*{Author Contributions.}
C.F-R.$^\dagger$, J.M.G-N.$^\dagger$, and R.F. conceptualized the study. C.F-R. and J.M.G-N. developed the clinical methodology, while R.F. developed the modeling and statistical methodology, performed the formal analysis and visualization, and wrote the necessary software code. Clinical investigation and data curation were performed primarily by C.F-R., I.G-C., and J.M.G-N. Project administration was handled by C.F-R.; resources were provided by S.H.F. and C.F-R. Clinical supervision was provided by S.H.F. and C.F-R.; analytical supervision was by R.F. Validation of the results involved C.F-R., J.M.G-N., S.H.F., and R.F. The original manuscript draft was written jointly by C.F-R.$^\dagger$, J.M.G-N.$^\dagger$, and R.F. All authors (C.F-R., J.M.G-N., I.G-C., S.H.F., R.F.) reviewed, edited, and approved the final manuscript. \\
$^\dagger$These authors contributed equally to this work.

\subsection*{Acknowledgements}
We are truly grateful to every patient who generously dedicated their time and effort to be part of our study. We also want to convey our deep appreciation to the exceptional medical and nursing staff of the University Hospital Nuestra Señora de Candelaria (Santa Cruz de Tenerife, Spain). These findings stem from the \textit{doctoral thesis} of Carmen Frias Ruiz, conducted at the University of Granada within the \textit{PhD Programme of Clinical Medicine and Public Health}, under the direction of Rene Fabregas.


 \section{Appendix.}
\subsection*{List of Abbreviations}

\begin{table}[htbp] 
    \caption{List of Abbreviations Used Throughout the Manuscript and Supplementary Information.}
    \label{tab:abbreviations}
    \centering 
    \footnotesize 

    \begin{minipage}[t]{0.48\textwidth} 
        \centering 
        \begin{tabular}{@{}ll@{}} 
            \toprule
            \textbf{Abbr.} & \textbf{Full Term} \\
            \midrule
            Adj.    & Adjusted                      \\
            Amb.    & Ambulatory                    \\
            AUC     & Area Under the Curve          \\
            CI      & Confidence Interval           \\
            Coeff.  & Coefficient                   \\
            \addlinespace 
            CRP     & C-reactive Protein            \\
            Ctr     & Center                        \\
            Diag:   & Diagnosis: (Prefix)           \\
            DM      & Diabetes Mellitus (in Diag:)  \\
            Infec.  & Infectious (in Diag:)         \\
            M       & Male                          \\
            \addlinespace
            MCV     & Mean Corpuscular Volume       \\
            OR      & Odds Ratio                    \\
            PEG     & Polyethylene Glycol           \\
            PR      & Prevalence Ratio              \\
            Ref.    & Reference                     \\
            \bottomrule
        \end{tabular}
    \end{minipage}
    \hfill 
    \begin{minipage}[t]{0.48\textwidth} 
        \centering 
        \begin{tabular}{@{}ll@{}} 
            \toprule
            \textbf{Abbr.} & \textbf{Full Term} \\
            \midrule
            Ext         & External                      \\
            F           & Female                        \\
            FOL         & Serum Folate                  \\
            GI/Neo      & Gastrointestinal/Neoplastic   \\ 
            Hb          & Hemoglobin                    \\
            \addlinespace
            Hema        & Hematologic (in Diag:)        \\
            Hepa        & Hepatic (in Diag:)            \\
            Hosp        & Hospital                      \\
            HTA         & Hypertension (in Diag:)       \\
            IQR         & Interquartile Range           \\
            Onco        & Oncologic (in Diag:)          \\
             \addlinespace
            RDW         & Red Cell Distribution Width   \\
            Renal       & Renal (in Diag:)              \\ 
            Rheum/AI    & Rheumatologic/Autoimmune      \\
            ROC         & Receiver Operating Char.      \\
            SD          & Standard Deviation            \\
            Sens        & Sensitivity                   \\
            Spec        & Specificity                   \\
            Susp.       & Suspicion                     \\
            Unspec.     & Unspecified                   \\
            VB12        & Vitamin B12                   \\
            VB12 Pre    & VB12 Pre-PEG                  \\
            VIP         & Variable Importance           \\
            vs.         & Versus                        \\
             \bottomrule
        \end{tabular}
    \end{minipage}
\end{table}

\newpage
\printbibliography
\end{document}


\maketitle



\section{Supplementary Methods.}
\textbf{Imputation.} Missing data for laboratory variables (CRP, Folate) were handled using Multiple Imputation by Chained Equations (MICE) with predictive mean matching, generating $m=5$ imputed datasets \cite{vanBuuren2010}. All subsequent analyses were performed on each imputed dataset, and results were pooled using Rubin's rules \cite{Rubin1987}.
\textbf{Beta Regression.} The outcome variable, PEG Recovery (\%), was transformed to the (0,1) interval using the standard transformation $(y \times (n-1) + 0.5) / n$, where $y$ is the recovery percentage divided by 100 and $n$ is the total sample size (n=875), to accommodate potential values of 0\% or 100\% while meeting the requirements for beta regression \cite{Smithson2006}. Beta regression models with a logit link function were fitted using the \texttt{betareg} package in R \cite{CribariNeto2004}. Model diagnostics included visual inspection of quantile residuals plotted against linear predictors.
\textbf{Random Forest.} Random Forest classification models were built using the \texttt{ranger} package in R \cite{Wright2017} to predict binary MacroB12 status. Models were trained on each imputed dataset using 500 trees, default settings for `mtry` (number of variables considered at each split, $\approx \sqrt{p}$), and a minimum terminal node size of 5. Variable importance was assessed using the permutation method, averaged across all imputations via the \texttt{vip} package.

\section{Supplementary Results.}
\subsection{Supplementary Data Characteristics.}

Missing data were present in several key laboratory variables prior to imputation. Supplementary Table~\ref{tab:missing_data} details the extent and pattern of missingness for all variables included in the primary and secondary analyses. The highest proportions of missing values were observed for C-reactive protein (CRP, approx. 50\%) and serum folate (approx. 21\%), with several other hematological parameters exhibiting moderate missingness (approx. 7\%). Demographic data, pre-PEG Vitamin B12, and diagnostic classifications generally had minimal missing data (<1\%). The patterns observed informed the selection of the multiple imputation strategy described in the main manuscript's Methods section (Section~\ref{sec:stats}).

\begin{table}[H] 
    \centering
    \sisetup{round-mode=places, round-precision=1, table-format=3.1}
    \caption{Summary of Missing Data for Key Analysis Variables (N=875).}
    \label{tab:missing_data}
    \begin{tabular}{@{}l r S[table-format=2.1] c@{}}
        \toprule
        Variable                       & N Missing & {\% Missing} & Type \\
        \midrule
        CRP (mg/L)                     & 435       & 49.7         & Numeric \\
        Serum Folate (ng/mL)           & 180       & 20.6         & Numeric \\
        Hemoglobin (g/dL)              & 63        & 7.2          & Numeric \\
        MCV (fL)                       & 63        & 7.2          & Numeric \\ 
        RDW (\%)                       & 63        & 7.2          & Numeric \\ 
        VB12 Pre-PEG (pg/mL)           & 1         & 0.1          & Numeric \\
        PEG Recovery (\%)              & 3         & 0.3          & Numeric \\
        \addlinespace
        Age (years)                    & 5         & 0.6          & Numeric \\
        Sex                            & 1         & 0.1          & Factor \\
        Source                         & 1         & 0.1          & Factor \\
         \addlinespace
        Diagnosis: Hematologic         & 2         & 0.2          & Factor (0/1) \\
        Diagnosis: Hepatic             & 2         & 0.2          & Factor (0/1) \\
        Diagnosis: Oncologic           & 2         & 0.2          & Factor (0/1) \\
        Diagnosis: Infectious          & 2         & 0.2          & Factor (0/1) \\
        Diagnosis: Renal               & 2         & 0.2          & Factor (0/1) \\
        Diagnosis: GI/Neo. Suspicion   & 2         & 0.2          & Factor (0/1) \\
        Diagnosis: Rheum./Autoimmune   & 1         & 0.1          & Factor (0/1) \\
        Diagnosis: Hypertension        & 0         & 0.0          & Factor (0/1) \\
        Diagnosis: Diabetes            & 0         & 0.0          & Factor (0/1) \\
        Diagnosis: Healthy/Unspecified & 0         & 0.0          & Factor (0/1) \\
        \midrule
        MacroB12 Status (Outcome)      & 0         & 0.0          & Factor \\
        \bottomrule
        \addlinespace[0.5ex]
    \end{tabular}
    \caption*{\footnotesize Missing data counts and percentages calculated prior to multiple imputation based on the total cohort size of N=875. Variables are grouped by type. Factor types for diagnoses reflect their binary (0/1) nature in the dataset used for imputation and modeling.}
\end{table}

\subsection{Minimum Detectable Effect Size (MEDS) Analysis.}

To provide context regarding the statistical sensitivity of our study given the available sample size (N=875), we estimated the minimum detectable effect size (MEDS) for key continuous and binary predictors. Calculations were performed based on the first imputed dataset, targeting 80\% statistical power at a two-sided significance level of \(\alpha=0.05\). The results (Supplementary Table~\ref{tab:meds_details}) indicate the magnitude of standardized effects (Cohen's d for continuous variables, Cohen's f\textsuperscript{2} for binary predictors in logistic regression approximations) that our study was well-powered to detect.

\begin{table}[htbp] 
    \centering
    \sisetup{round-mode=places, round-precision=3, table-format=1.3, table-number-alignment = center}
    \caption{Minimum Detectable Effect Sizes (MEDS) for Selected Variables (\(\alpha=0.05\), Power=0.80, N=875).}
    \label{tab:meds_details}
    \footnotesize 
    \begin{tabular}{@{} p{5cm} l S[table-format=1.3] p{2.5cm} p{2.8cm} @{}}
        \toprule
        Variable                       & Test Type             & {MEDS (Std.)} & MEDS (Approx. Orig. Units) & Calculation Details \\
        \midrule
        \multicolumn{5}{l}{\textit{Continuous Variables (Compared by MacroB12 Status)}} \\
        \addlinespace[0.5ex]
        Age (years)                    & T-test (Means)        & 0.213         & 4.5 years                  & d; SD\textsubscript{p}=21.1 \\ 
        VB12 Pre-PEG (pg/mL)           & T-test (Means)        & 0.213         & 287.2 pg/mL                & d; SD\textsubscript{p}=1346.1 \\ 
        Serum Folate (ng/mL)           & T-test (Means)        & 0.213         & 1.0 ng/mL                  & d; SD\textsubscript{p}=4.6 \\ 
        Hemoglobin (g/dL)              & T-test (Means)        & 0.213         & 0.5 g/dL                   & d; SD\textsubscript{p}=2.4 \\ 
        CRP (mg/L)                     & T-test (Means)        & 0.213         & 1.4 mg/L                   & d; SD\textsubscript{p}=6.4 \\ 
        MCV (fL)                       & T-test (Means)        & 0.213         & 1.5 fL                     & d; SD\textsubscript{p}=7.2 \\ 
        RDW (\%)                       & T-test (Means)        & 0.213         & 0.5 \%                     & d; SD\textsubscript{p}=2.5 \\ 
        \addlinespace
        \multicolumn{5}{l}{\textit{Binary Predictors (Logistic Regression Framework Approximation)}} \\
        \addlinespace[0.5ex]
        Sex (Male vs Female)           & LogReg (Bin. Pred.)   & 0.0091        & N/A                        & f\textsuperscript{2}; R\textsuperscript{2}=0.047 \\ 
        Diagnosis: Rheum./Autoimmune   & LogReg (Bin. Pred.)   & 0.0091        & N/A                        & f\textsuperscript{2}; R\textsuperscript{2}=0.030 \\ 
        Diagnosis: Hematologic         & LogReg (Bin. Pred.)   & 0.0091        & N/A                        & f\textsuperscript{2}; R\textsuperscript{2}=0.088 \\ 
        Diagnosis: Oncologic           & LogReg (Bin. Pred.)   & 0.0091        & N/A                        & f\textsuperscript{2}; R\textsuperscript{2}=0.044 \\ 
        \bottomrule
        \addlinespace[0.5ex]
    \end{tabular}
    \caption*{\footnotesize MEDS calculations based on imputed dataset \#1 (N=875). Continuous variable MEDS based on \texttt{pwr::pwr.t2n.test} solving for Cohen's d. Binary predictor MEDS based on \texttt{pwr::pwr.f2.test} solving for Cohen's f\textsuperscript{2}, approximating logistic regression power. R\textsuperscript{2} in details refers to the proportion of variance in the binary predictor explained by other predictors in the model. N/A: Not Applicable (direct conversion to original units is not straightforward for f\textsuperscript{2}). \textbf{Abbreviations:} d, Cohen's d; SD\textsubscript{p}, Pooled Standard Deviation; f\textsuperscript{2}, Cohen's f-squared.}
\end{table}

\subsection{Discriminatory Performance of Secondary Variables.}
\begin{figure}[ht!]
    \centering
    \includegraphics[width=\textwidth]{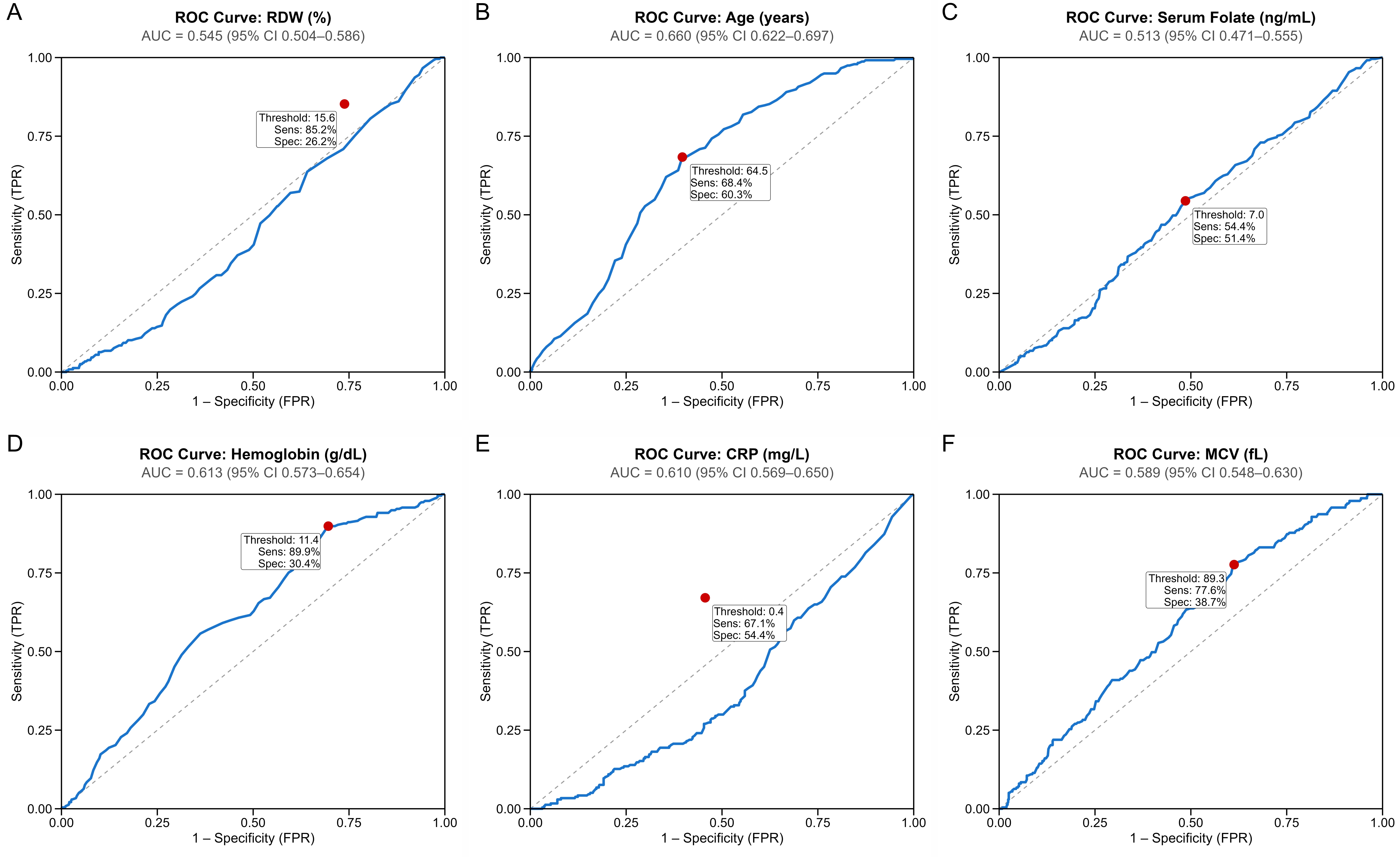}
    \caption{\textbf{Supplementary ROC Curve Analyses.} ROC curves illustrating the discriminatory ability for MacroB12 positive status for: \textbf{(A)} RDW (\%), \textbf{(B)} Age (years), \textbf{(C)} Serum Folate (ng/mL), \textbf{(D)} Hemoglobin (g/dL), \textbf{(E)} CRP (mg/L), and \textbf{(F)} MCV (fL). The Area Under the Curve (AUC) and 95\% confidence interval (CI) are indicated in each panel title. Red circles mark the optimal threshold determined by the Youden index, with corresponding performance metrics annotated. The diagonal dashed line represents random chance (AUC = 0.5).}
    \label{fig:roc_supp}
\end{figure}

While pre-PEG Vitamin B12 demonstrated the best individual performance for discriminating MacroB12 status (see Main Text, Figure~\ref{fig:roc_main}), we also evaluated other continuous variables using ROC analysis (Supplementary Fig.~\ref{fig:roc_supp}). Increased patient age showed a statistically significant but only modest ability to discriminate (Area Under the Curve [AUC] = 0.660, 95\% CI 0.622--0.697), with an optimal Youden index threshold at 64.5 years yielding a sensitivity of 68.4\% and specificity of 60.3\% (Supplementary Fig.~\ref{fig:roc_supp}B). Hemoglobin levels provided poor discrimination (AUC = 0.613, 95\% CI 0.573--0.654), achieving high sensitivity (89.9\%) only at the cost of very low specificity (30.4\%) at its optimal threshold of 11.4 g/dL (Supplementary Fig.~\ref{fig:roc_supp}D). Similarly, CRP (AUC = 0.610, 95\% CI 0.569--0.650; Optimal Threshold 0.4 mg/L) and MCV (AUC = 0.589, 95\% CI 0.548--0.630; Optimal Threshold 89.3 fL) offered statistically significant but poor overall discrimination (Supplementary Figs.~\ref{fig:roc_supp}E, F). RDW showed marginal significance and very poor performance (AUC = 0.545, 95\% CI 0.504--0.586) (Supplementary Fig.~\ref{fig:roc_supp}A). Serum folate was not a significant discriminator (AUC = 0.513, 95\% CI 0.471--0.555) (Supplementary Fig.~\ref{fig:roc_supp}C). These findings underscore the limited utility of these individual markers compared to pre-PEG B12 for identifying potential MacroB12 positivity.

\subsection{Factors Associated with PEG Recovery Percentage.}
To elucidate factors potentially influencing the PEG precipitation assay itself, we modelled the quantitative \% Recovery using Beta regression on pooled imputed data. The multivariable analysis, adjusting for the same covariates as in the main MacroB12 status models, identified several independent associations (Supplementary Fig.~\ref{fig:supp_beta_rf}A). Higher pre-PEG Vitamin B12 concentration (\textit{P}<0.001), older age (\textit{P}=0.001), and higher MCV (\textit{P}=0.001) were strongly associated with a lower mean \% Recovery, suggesting that patient characteristics associated with MacroB12 positivity also influence the degree of precipitation observed. Conversely, elevated CRP levels (\textit{P}=0.03) and specific clinical contexts, namely Oncologic (\textit{P}=0.003) or Healthy/Unspecified (\textit{P}=0.03) diagnoses, were associated with significantly higher mean \% Recovery compared to other diagnostic groups. The biological or analytical basis for these latter associations warrants further investigation. Diagnostic evaluation via quantile residuals indicated the Beta regression model provided a reasonable fit to the data distribution (Supplementary Fig.~\ref{fig:supp_beta_rf}B), substantially improving upon standard linear model assumptions observed with simple log-transformation (data not shown).

\subsection{Predictive Variable Importance via Random Forest.}
\begin{figure}[ht!]
    \centering
    \includegraphics[width=\textwidth]{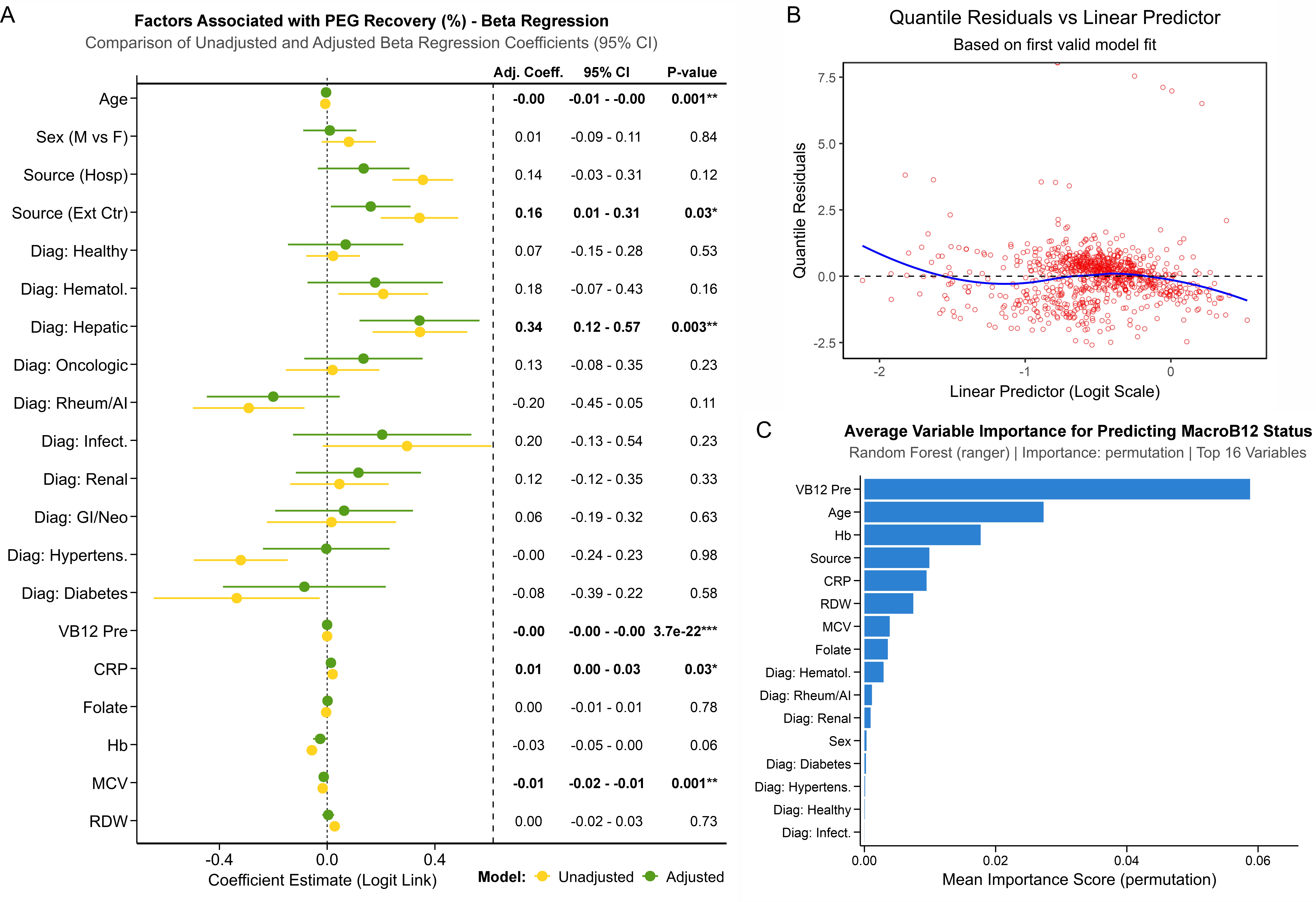}
    \caption{\textbf{Supplementary Analyses: Beta Regression for PEG Recovery and Random Forest Variable Importance.}
    \textbf{(A)} Forest plot comparing unadjusted (yellow) and adjusted (green) coefficients from Beta regression models predicting PEG Recovery (\%). Coefficients represent change on the logit scale. Horizontal lines indicate 95\% confidence intervals (CIs). Numerical values correspond to the adjusted coefficient, 95\% CI, and P-value (\textsuperscript{***}P < 0.001, \textsuperscript{**}P < 0.01, \textsuperscript{*}P < 0.05).
    \textbf{(B)} Diagnostic plot showing quantile residuals versus the linear predictor (logit scale) from the adjusted Beta regression model, based on the first valid imputed dataset fit. The blue line represents a LOESS smooth.
    \textbf{(C)} Average variable importance scores derived from Random Forest models trained on multiply imputed datasets to predict MacroB12 status. Importance was calculated using the permutation method, averaged across imputations. Variables are ranked by mean importance.}
    \label{fig:supp_beta_rf}
\end{figure}

Complementing the association analyses, we employed a Random Forest machine learning model to assess the relative importance of all variables in jointly predicting MacroB12 status, capturing potential non-linearities and interactions. Variable importance was quantified using the mean decrease in model accuracy upon permutation, averaged across imputed datasets. Consistent with the regression model findings (Results Section \ref{sec:results}), pre-PEG Vitamin B12 emerged as the variable with the highest predictive importance, followed by patient age and hemoglobin level (Supplementary Fig.~\ref{fig:supp_beta_rf}C). Several other variables, including patient source, CRP, RDW, and MCV, also demonstrated non-negligible importance scores, suggesting they contribute collectively to the predictive signal, even if their individual adjusted associations were not statistically significant in the regression models. Most diagnostic categories exhibited relatively low importance in this predictive framework. These results reinforce the primary roles of pre-PEG B12 and age in relation to MacroB12 status within this cohort.

\subsection{Sensitivity Analysis: Impact of a Different MacroB12 Threshold.}
\label{sec:supp_sensitivity_threshold}

The primary analyses presented in the main manuscript utilized a polyethylene glycol (PEG) recovery threshold of <30\% to define MacroB12 positivity. Given that reported thresholds for PEG-based MacroB12 detection can vary in the literature, often cited within the range of <30\% to <40\% \cite{Delgado2024, Remacha2014, Soleimani2019}, we conducted a sensitivity analysis to evaluate the robustness of our principal findings to an alternative, slightly less stringent definition of MacroB12 positivity using a PEG recovery threshold of <40\%.

For this sensitivity analysis, MacroB12 status was re-classified based on \% Recovery <40\% using the same imputed datasets (N=875 from the primary cohort derivation; see Main Text, Figure~\ref{fig:study_flow}). Under this alternative definition, the prevalence of MacroB12 positivity increased to 50.9\% (n=445 MacroB12 Positive; n=430 MacroB12 Negative), compared to 27.1\% with the <30\% threshold.

The fully adjusted multivariable logistic regression model was re-fitted using this <40\% recovery threshold to define the outcome. Key predictors identified in the primary analysis were re-evaluated (Supplementary Table~\ref{tab:supp_sensitivity_or}). Older age (aOR 1.01, 95\% CI 1.01--1.02; \textit{P}=5.60$\times$10\textsuperscript{-4}) and pre-PEG Vitamin B12 concentration (aOR 1.00, 95\% CI 1.00--1.00; \textit{P}=1.65$\times$10\textsuperscript{-16}) remained strongly and significantly associated with increased odds of MacroB12 positivity. A Rheumatologic/Autoimmune diagnosis also retained its significant association (aOR 2.50, 95\% CI 1.08--5.78; \textit{P}=0.032). Notably, under this broader definition, a diagnosis of Diabetes Mellitus emerged as a significant predictor (aOR 4.53, 95\% CI 1.32--15.5; \textit{P}=0.017). However, the paradoxical associations for hemoglobin (aOR 1.09, 95\% CI 0.999--1.19; \textit{P}=0.054) and MCV (aOR 1.02, 95\% CI 0.996--1.05; \textit{P}=0.095) were attenuated and no longer met statistical significance. The protective association of a Hematologic diagnosis also lost significance (aOR 0.80, 95\% CI 0.34--1.87; \textit{P}=0.606).

Furthermore, the discriminatory performance of pre-PEG Vitamin B12 concentration for identifying MacroB12 positivity (defined as <40\% recovery) was re-assessed via ROC curve analysis. The Area Under the Curve (AUC) was **0.713 (95\% CI 0.679--0.746)**, which is marginally lower than the AUC of 0.744 observed with the <30\% threshold (Main Text, Figure~\ref{fig:roc_main}). The optimal Youden-indexed threshold for pre-PEG B12 under this alternative MacroB12 definition was **1579.5 pg/mL**, yielding a sensitivity of **58.2\%** and a specificity of **75.3\%**.

Collectively, these sensitivity analyses indicate that while adopting a <40\% PEG recovery cutoff increases the apparent prevalence of MacroB12 and alters the statistical significance of some secondary predictors (notably Hb, MCV, Hematologic Dx losing significance, and Diabetes Dx gaining significance), the core associations for pre-PEG B12, age, and Rheumatologic/Autoimmune diagnosis remain robust. The discriminatory performance of pre-PEG B12, with an AUC of 0.713, is marginally reduced compared to the primary analysis but still indicative of moderate utility. This supports the generalizability of our primary conclusions regarding key factors, while highlighting that precise effect estimates and the significance of less dominant predictors can be sensitive to the specific diagnostic threshold employed for MacroB12.

\begin{table}[H]
    \centering
    \caption{Comparison of Multivariable-adjusted OR for Key Predictors of MacroB12 Positivity using Alternative PEG Recovery Thresholds.}
    \label{tab:supp_sensitivity_or}
    \footnotesize
    \begin{tabular}{@{}l c c@{}}
        \toprule
        Predictor Variable & aOR (95\% CI) with <30\% Threshold & aOR (95\% CI) with <40\% Threshold \\
        & (Primary Analysis) & (Sensitivity Analysis) \\
        \midrule
        Age (per year)                  & 1.03 (1.02--1.04)\textsuperscript{***} & 1.01 (1.01--1.02)\textsuperscript{***} \\
        VB12 Pre-PEG (per pg/mL)        & 1.00 (1.00--1.00)\textsuperscript{***} & 1.00 (1.00--1.00)\textsuperscript{***} \\
        Hemoglobin (per g/dL)           & 1.14 (1.02--1.27)\textsuperscript{*}   & 1.09 (0.999--1.19) \\
        MCV (per fL)                    & 1.04 (1.01--1.07)\textsuperscript{*}   & 1.02 (0.996--1.05) \\
        Diagnosis: Rheum./Autoimmune    & 2.96 (1.24--7.10)\textsuperscript{*}   & 2.50 (1.08--5.78)\textsuperscript{*} \\
        Diagnosis: Hematologic          & 0.32 (0.10--0.97)\textsuperscript{*}   & 0.80 (0.34--1.87) \\
        Diagnosis: Diabetes             & 1.98 (0.70--5.58)                      & 4.53 (1.32--15.5)\textsuperscript{*} \\
        \bottomrule
    \end{tabular}
    \caption*{\footnotesize aOR: Multivariable-adjusted Odds Ratio; CI: Confidence Interval. Models adjusted for all covariates listed in main text Methods. \textsuperscript{***}P<0.001, \textsuperscript{**}P<0.01, \textsuperscript{*}P<0.05. N=875. Values for <30\% threshold are from the multivariable logistic regression model presented in Figure~\ref{fig2:forest_plots}B in the main manuscript. Results for <40\% threshold derived from re-analysis.}
\end{table}

\printbibliography